\documentclass[a4paper,11pt]{article}
\pdfoutput=1 % if your are submitting a pdflatex (i.e. if you have
\usepackage{jcappub} % for details on the use of the package, please
\usepackage[T1]{fontenc} % if needed
\RequirePackage{bm,amsfonts,lineno,amsmath,amssymb,float,eurosym,latexsym,epsf,mathtools,cuted,makecell,array}
\usepackage[font={footnotesize,it}]{caption}
\usepackage{epstopdf}
\setcellgapes{4pt}%parameter for the spacing

\title{\boldmath Relativistic charged spheres: Compact stars, compactness and stable configurations}

\author[a]{J. Kumar}
\author[b,1]{S. K. Maurya\note{Corresponding author} }
\author[a]{A. K. Prasad}
\author[c]{Ayan Banerjee}

\affiliation[a]{Department of Applied Mathematics, Central University of Jharkhand, Ranchi-835205, India}
\affiliation[b]{Department of Mathematical and Physical Sciences,
College of Arts and Science, University of Nizwa, Nizwa, Sultanate of Oman}
\affiliation[c]{Astrophysics and Cosmology Research Unit, University of KwaZulu Natal, Private Bag X54001, Durban 4000,
South Africa}

\emailAdd{jitendark@gmail.com}
\emailAdd{sunil@unizwa.edu.om}
\emailAdd{amitkarun5@gmail.com}
\emailAdd{ayan\_7575@yahoo.co.in}

\abstract{This paper aims to explore a class of static stellar equilibrium configuration of relativistic charged spheres made of a charged perfect fluid.  Solving the Einstein-Maxwell field equations, we consider a particularized metric potential, Buchdahl ansatz [ Phys. Rev.116, 1027 (1959)] and then by using a simple transformation. The study is developed by matching the interior region with Riessner-Nordstr$\dot{\text{o}}$m metric as an exterior solution. The matter content the charged sphere satisfies all the energy conditions and hydrostatic equilibrium equation, i.e. the modified Tolman-Oppenheimer-Volkoff (TOV) equation for the charged case is maintained. In addition to this, we also discuss some important properties of the charged sphere such as total electric charge, mass-radius relation, surface redshift, and the speed of sound are analyzed. Obtained solutions are presented by the graphical representation that provides strong evidence for a more realistic and viable stellar structure. Obtained results are compared with analogue objects with similar mass and radii, such as SAX J1808.4-3658, 4U 1538-52, PSR J1903+327, Vela X-1, and 4U1608-52. It is also noted that the Buchdahl ansatz for a given transformation provides a physically viable solution only for the charged case when $0 < K < 1$, where density and pressure are maximum at the center and monotonically decreasing towards the boundary. Obtained results are also quite important both from theoretical and astrophysical scale to analyze other compact objects such as white dwarfs, neutron stars, boson stars, and  quark stars.}   


\begin{document}
\maketitle
\flushbottom

\section{Introduction} \label{sec:intro}
Compact astrophysical objects, such as neutron stars (NSs) and quark stars (QSs), are become an excellent test-bed to probe certain properties of gravitational fields. Particularly, the relativistic stellar models have been studied for a couple of decades, though the first exact solution of Einstein's field equation was obtained by Schwarzschild in 1916,  for the interior of a compact object in hydrostatic equilibrium. This leads as a starting point to the learning of an exact solution for stellar object satisfying a variety of criteria which is physical admissible. Moreover, exact solution plays an important role in development of many areas of  gravitational field such as  solar system test, black hole solution, stellar modelling, gravitational collapse and so on. In fact, obtaining a singularity free interior solution for compact astrophysical objects  have an important consequences when it comes to solving the field equations. Thus, studying compact objects from their microscopic composition and properties of dense matter is one of the most fundamental problems in modern astrophysics. This implies not only the search of new solutions and material composition, but also the research for some special properties, related to their structure, in comparison with observational data.

 Recently it has been  suggested from observations that relativistic compact stars may soon provide information about supra-nuclear equation of state (EoS), i.e. the relation between internal density and pressure at densities beyond nuclear \cite{Lattimer,Lattimer1}. Thus, it is important to measure the mass-radius $M-R$ relation
 \cite{Burikham,Boehmer1,Chandrasekhar,Herrera1}  of compact objects which is highly sensitive to the EoS \cite{Ray,Negreiros,Varela,Maharaj1,Ivanov1}, because the interior structure of such compact stars can vary with its mass. When the mass of a star is sufficiently large, its central density becomes more higher, and thus the possibility of exotica in their inner cores which cause variability in their exterior gravitational field. This statement sometimes refer  to produce a gravitational field different from that of stars without such exotica in their inner cores. In spite of this reasoning and  lack of understanding about such compact objects, the motivation was initiated by the discovery of the pulsar PSR J1614-2230 yielded a mass of  1.97 $\pm$ 0.04 $M_{\odot}$. Its very high inclination, at ${89}^{\circ}$.17, allowed the detection of a strong Shapiro delay signature by an X-ray telescope, like NICER \cite{Gendreau}. In particular, this high mass of pulsar provides a lower limit on the maximum mass of neutron stars and fix a dividing line between black holes and neutron stars to at least this value. There has been an-ongoing effort to understand many astrophysical compact objects whose estimated masses and radii are not compatible with our known sources, such as X-ray pulsar Her X-1, X-ray burster 4U 1820-30, X-ray sources 4U 1728-34, PSR 0943+10 and RX J185635-3754. Thus the aim of this study is the prediction of masses and radii of compact objects which could  constrain the EoS of matter in the high density regime. Many authors discuss about mass-radius of compact objects (see more and further references \cite{E,R,F,Guver}). Similar types of models have also been extended in different analogy and context  \cite{Jose1,Jose2,Maurya3,Maharaj2,Barreto1,Barreto2,Herrera2,Maharaj3,Maharaj4}.

The theoretical possibility of studying self-gravitating fluid models  with an effect of electric charge and electric field  has been done previously by different authors. As in evidence Rosseland \cite{Rosseland} (see also Eddington \cite{Eddington}), studied the possibility of a self gravitating star treated as a ball of hot ionized gas containing a certain amount of charge on Eddington's theory.  In such a system large number of electrons (as compared to positive ions) run to escape from its surface due to their higher kinetic energy and the motion of electrons will continue until the electric field induced inside the star prevents more electrons to escape from its surface. In this way, a star would contain only positive ions.  Later on, it was proved that equilibrium is attained after some amount of electrons escape and the net electric charge about 100 Coulombs per solar mass. At this point of view collapsing of a star to a point singularity may be avoided by the effects of  charge as the gravitational attraction will then be counter balanced by the Coulomb repulsion. Another point is that the electromagnetic interaction between particles is stronger than the gravitational interaction by a factor of $\lambda_G= e^2/Gm_u^2 \approx 1.25 \times 10^{36}$ where the proton charge is denoted by $e$ and $m_u$ = 931.5 MeV/c$^2$ is the atomic mass unit, with $c$ being the speed of light.  With this  assumption of the of electrostatic interactions, one can argue that gravitational forces should play the dominant role in determining stellar structure \cite{Krivoruchenko}.
Bekenstein \cite{Bekenstein} first analyses the stability of charged fluid spheres by generalizing the Oppenheimer-Volkoff and then by other authors in \cite{Zhang,Felice,Yu,Felice1}. Physical behaviour and stability of charged dust stars was discovered by Majumdar \cite{Majumdar},  Papapetrou \cite{Papapetrou}, Bonnor \cite{Bonnor} and by Skenderis \cite{skenderis1,skenderis2} to name a few. Some studies have also concluded that due to presence of electromagnetic field that affects the value of luminosities, redshifts, and maximum mass of a compact relativistic object 
(see \cite{Ivanov,S1} for details).

The search for exact solutions of Einstein-Maxwell field equations for static spherically symmetric metric with isotropic matter is of continuous interest to researcher. Although a lot of solutions have been proposed by several authors and their observational implications along with theoretical structures have been studied
(check \cite{Bhar,Bhar1,Takisa,Maurya1,Maurya2,Maurya4,Lemos,Kouretsis,Rahaman,Hansraj,Gupta4,Esculpi,jk,jitendra} and references therein). The exact Einstein-Maxwell solution can be obtained by specifying a particular form of gravitational potentials with linear EoS consistent with quark matter in \cite{Komathiraj}. Also, Varela \textit{et al} \cite{Varela} have found a new solution for charged anisotropic matter with linear or nonlinear EoS.  Consequently, many simplifying assumptions have been taken in order to integrate the field equations, one of them is the assumption of metric potentials due to any reliable information of an  EoS at extreme densities. One of them is the Vaidya-Tikekar ansatz \cite{Vaidya}, and a large number of solutions have been studied  to generate and analyze physically viable models of compact astrophysical objects \cite{10,11,12,Kumar1}.

There is another type of metric ansatz due to Buchdahl \cite{Buchdahl} who proposed an important scheme for finding physically reasonable spherically symmetric perfect fluid solution, which possesses monotonically decreasing density towards the boundary. After that Vaidya and Tikekar  \cite{Vaidya} particularized Buchdahl ansatz by giving a geometry for the interior physical 3-space of the configuration. In this connection, Khugaev \textit{et al} \cite{Khugaev} have obtained higher dimensional solutions for super compact star by utilize the ansatz of Buchdahl-Vaidya-Tikekar and extended their work in pure Lovelock garvity \cite{Molina}. The present work, we shall utilize the Buchdahl \cite{Buchdahl} ansatz as a metric potential and introduce a well known transformation to determine the unknown variables that describe the interior of a stellar configuration.

The paper has been organized as follows:  following a brief introduction in Sec. \ref{sec:intro},  the relevant Einstein-Maxwell system of equations describing a charged relativistic stellar configuration has been laid down in Sec. \ref{cim}.  Then we have solved the system of equations by paying a particular attention on the metric potential, namely,  Buchdahl \cite{Buchdahl} in the same section. In Sec. \ref{bou}, we matched the interior solution to an exterior Reissner-Nordstr$\dot{{o}}$m line element and determine the constant coefficient. Next, in Sec. \ref{phy}, some physical features of the model have been discussed in briefly and obtained results are compared with Observational stellar mass data. Finally in Sec. \ref{fin}, we  have concluded by highlighting some features of our model.

 \section{Charged isotropic matter} \label{cim}
With the purpose of analyzing the properties of  relativistic charged fluid distribution,
 we assume the line element in Schwarzschild coordinates, as
 \begin{eqnarray}
 ds^{2}= e^{\nu(r)}dt^{2}-e^{\lambda(r)}dr^{2} -r^{2}(d\theta^{2}+\sin^{2}\theta d\phi^{2}),\label{1}
\end{eqnarray}
where $\nu(r)$ and $\lambda(r)$ are arbitrary functions of the radial
coordinate $r$, which yet to be determined by solving the field equations.

We are interested in Einstein- Maxwell equations in the presence of charged matter.
The properties of the stellar matter can be understood by including the terms from the Maxwell's equation, which is written as $T^{i}_{j}$= $M^{i}_{j}$+$E^{i}_{j}$. Here,
$M^{i}_{j}$ stands for the energy-momentum tensor of a perfect fluid and $E^{i}_{j}$ is the electromagnetic energy-momentum tensor. So, the complete form of Einstein-Maxwell field equations
 for a charged fluid sphere is defined as
 \begin{eqnarray} R^{i}_{j}-\frac{1}{2}R\delta^{i}_{j}=-\kappa T^{i}_{j} =-\kappa \big[(c^{2}\rho+p)u^{i}u_{j}-p\delta^{i}_{j}+\frac{1}{4\pi}(-F^{im}F_{jm}
 +\frac{1}{4}\delta^{i}_{j}F_{mn}F^{mn})\big],\label{2}
\end{eqnarray}
with $ \kappa=8\pi G/c^{4}$, while $\rho$ is the matter density and $p$ is the pressure which is measured
relative to the comoving fluid 4-velocity $u^i = e^{-\nu}\delta^{i}_{0}$. We assume that
the interior of the star is filled with a perfect fluid and the form of energy-momentum tensor is written in
the right hand side of Eq. (\ref{2}), which is $\left[(c^{2}\rho+p)u^{i}u_{j}-p\delta^{i}_{j}\right]$.
One of the most important argument to assume a perfect fluid is that
the flow of matter is adiabatic, no heat conduction or viscosity is present \cite{Misner}.
The second term $E^{i}_{j}$ is the part of energy momentum tensor due to electromagnetic
fields and is defined by $\left(-F^{im}F_{jm}+{1}/{4}\delta^{i}_{j}F_{mn}F^{mn}\right)/{4\pi}$,
where the electromagnetic tensor $F_{ij}$ satisfies Mexwells equations
\begin{eqnarray}
 F_{ik},j+F_{kj},i+F_{ji},k=0,\nonumber \\
 \left[\sqrt{-g}F^{i j}\right]_{,j}=4\pi J^{i}\sqrt{-g},\label{3}
 \end{eqnarray}
where $J^{i}$ is the electric current density is written as $J^{i}= \sigma u^{i}$
with $F_{ij}$ denote the skew symmetric electromagnetic field tensor. Imposing that there is a
static spherically symmetric electric field, the only non-vanishing components $F^{01}=- F^{10}$,
being a function of the radial coordinate r alone, and the other terms are absent. Hence, from
Eq. (\ref{3}), we can obtain the following expression for non-vanishing component
\begin{eqnarray}
 q(r)= r^{2}\sqrt{-F_{01}F^{10}}=r^{2}F^{10}e^{(\lambda+\nu)/2} = 4\pi\int_{0}^{r}\sigma r^{2}e^{\lambda/2}dr,\label{4}
 \end{eqnarray}
where $q(r)$ is the total electric charge inside a sphere of radial coordinate $r$, which does
not depend on the timelike coordinate t with where $\sigma$ is the charge density. With the followings, $q(r)$ is
invariant under the transformation $q(r) = - q(r)$ and $\sigma = -\sigma$.

In the present case, with the metric (\ref{1}) and energy momentum tensor (\ref{2}),
the nonzero components of the Einstein equations provide the following relationships
 \begin{eqnarray}
 \frac{\lambda'}{r}e^{-\lambda}+\frac{(1-e^{-\lambda})}{r^{2}}=\kappa \rho(r)+\frac{q^{2}(r)}{r^{4}},\label{5}\\
\frac{\nu'}{r}e^{-\lambda}-\frac{(1-e^{-\lambda})}{r^{2}}=\kappa p(r)-\frac{q^{2}(r)}{r^{4}},\label{6} \\
\bigg(\frac{\nu''}{2}-\frac{\lambda'\nu'}{4}+\frac{\nu^{'2}}{4}+\frac{\nu'-\lambda'}{2r}\bigg)e^{-\lambda}=\kappa p(r)+\frac{q^{2}(r)}{r^{4}},\label{7}
\end{eqnarray}
with $ \kappa = 8\pi G/c^{4}$, and prime denotes the differentiation with respect to the radial
coordinate r. The system of Eqs. (\ref{5})-(\ref{7}) determines the behaviour of gravity for a charged perfect fluid. As usual, when $q(r )= 0$, one can restore the Einsteins equations for a perfect fluid. Here we are dealing with five unknowns $ \nu, \lambda, \rho( r), p(r)$ and $q(r)$, which we are going to solve to get our desire results. At this point, it is important to have an ansatz specifying one of the metric functions or select an EoS that's relate pressure and density which leads to analytical solutions.

The analysis presented in this article by considering a sensible choices for an ansatz of the metric potential due to Buchdahl \cite{Buchdahl}. For the metric function $e^{\lambda}$ we make the choice
\begin{eqnarray}
e^{\lambda}=\frac{K(1+Cr^{2})}{K+Cr^{2}},  ~~~{with}~~ C>0\label{8}
 \end{eqnarray}
where $K$ is an arbitrary constant. Using the Buchdahl ansatz for metric
potential is not new. The model, though exceedingly simple, satisfies
the physical constraints of a realistic star ensures the regularity
and finite conditions at the centre of the sphere. When $C =-K/R^2$, we
gain the metric function considered by Vaidya and Tikekar \cite{Vaidya}. This facilitated the model in an interesting geometric meaning as deviation from specificity of 3-space geometry.

Now, introducing a new coordinate transformation $e^{\nu}=Z^{2}(r )$, the field
Eqs. (\ref{5})-(\ref{7}) take the following form
 \begin{eqnarray}
\frac{K+Cr^{2}}{K(1+Cr^{2})}\Bigg[\frac{Z''}{Z}-\frac{Z'}{rZ}+\frac{C(K-1)r(Cr-Z'/Z)}{(K+Cr^{2})(1+Cr^{2})}\Bigg]
=\frac{2q^{2}(r)}{r^{4}}\label{9}
  \end{eqnarray}
Furthermore, at this stage it is convenient to introduce the following transformation (Gupta-Jasim \cite{Gupta2004} two step method) 
 \begin{eqnarray}
 X=\sqrt{\frac{K+Cr^{2}}{1-K}},~~~~and~~~~ Z=(1+X^{2})^{1/4}Y,\label{10}
 \end{eqnarray}
 Going back to the (\ref{9}) and using the transformation (\ref{10}) enable us to
 rewrite the second order differential equation in a simpler form, which is
 \begin{eqnarray}
\frac{d^{2}Y}{dX^{2}}+\psi Y=0,\label{11}
 \end{eqnarray}
where for notational simplicity we use
 \begin{eqnarray} \psi=-\frac{1}{(1+X^{2})}\Bigg[1-K+2K q^{2}\frac{1+Cr^{2}}{C^{2}r^{6}}+\frac{3X^{2}-2}{4(1+X^{2})}\Bigg]. \label{12}
 \end{eqnarray}
 To gain some more insight and solve the Eq. (\ref{11}) easily, we set $\psi$ as
\begin{eqnarray}
\psi=-\frac{2\alpha^{2}}{(X^{2}\alpha^{2}+X)},\label{13}
\end{eqnarray}
where $\alpha $ is a positive constant. Now the relation (\ref{12}) and (\ref{13})
leads to defining the total charge of the system as
\begin{eqnarray}
\frac{q^{2}}{r^{4}}=\frac{C^{2}r^{2}\bigg(8\alpha^{2}+4KX(1+X^{2})(1+X\alpha^{2})
-2X(1-7X\alpha^{2})-X^{3}(7-X\alpha^{2})\bigg)}{8K(1+Cr^{2})(X^{2}\alpha^{2}+X)}.\label{14}
\end{eqnarray}
This brings out a new form of equation when replacing Eq. (\ref{13}) into Eq. (\ref{11}) it gives
\begin{eqnarray}
 (X^{2}\alpha^{2}+X)\frac{d^{2}Y}{dX^{2}}-2\alpha^{2}Y=0. \label{15}
\end{eqnarray}
We now consider the simple derivation of the differential equation (\ref{15})
leads to the following expression for $Y$ (for further details see appendix A), as
\begin{eqnarray}
  Y(X)=(X^{2}\,\alpha^{2}+X)\,A\,\alpha^{2}\,\bigg[\,\sin^{2}(\arctan(\sqrt{X\alpha^{2}}\,))/2 -\csc^{2}(\arctan(\sqrt{X\alpha^{2}}\,))/2 \bigg] \nonumber \\ -(X^{2}\,\alpha^{2}+X)\,\bigg[\,A\,\alpha^{2}\,
\log(\sin^{2}(\arctan(\sqrt{X\alpha^{2}}\,))- B\,\bigg] \label{16}
\end{eqnarray}
where A and B are arbitrary integrating constant.
Plugging the values of (\ref{16}) into the Eq. (\ref{10}) for $Z$ reads as
\begin{eqnarray}
Z=(1+X^{2})^{1/4}\Bigg[(X^{2}\alpha^{2}+X)A\alpha^{2}H(X)+B(X^{2}\alpha^{2}+X)\Bigg],\label{17}
\end{eqnarray}
where $ H(X)=\Bigg(\sin^{2}(\arctan\sqrt{X\alpha^{2}})/2-\csc^{2}(\arctan\sqrt{X\alpha^{2}})/2
-\log\sin^{2}(\arctan\sqrt{X\alpha^{2}})\Bigg).$
Finally, the complete solution of the Einstein-Maxwell system (\ref{5})-(\ref{7}) is then given by
\begin{eqnarray}
 \kappa \rho(r)= -\frac{C^{2}r^{2}\psi}{8K(1+Cr^{2})^{2}(1-K)} \bigg[4(1+Cr^{2})^{2}+(1-K)(2+7Cr^{2}+K(1-4Cr^{2}))\bigg]\nonumber\\
 +\frac{C(K-1)(3+Cr^{2})}{K(1+Cr^{2})^{2}},~~\label{18}
 \end{eqnarray}

 \begin{eqnarray}
 \kappa p(r) = -\frac{C(K-1)}{K(1+Cr^{2})}+\frac{C^{2}r^{2}\psi}{8K(1+Cr^{2})^{2}(1-K)}\bigg[4(1+Cr^{2})+(1-K)(2+7Cr^{2}+K(1-4 Cr^{2}))\bigg]\nonumber \\ +\left[\frac{A1\times A2+A3\times A4}{A5\times A2}\right]\left(\frac{K+Cr^{2}}{K(1+Cr^{2})}\right) ,~~~~~~\label{19}
  \end{eqnarray}

 where we use the notations \\

$ A1=\frac{2C}{\sqrt{(1-K)(K+Cr^{2})}}\left[\frac{\left(\Pi\right)^{3/2}\alpha^{2}+\Pi}{2\left(\Upsilon\right)^{3/4}}+\left(\Upsilon\right)^{1/4}\left(2\sqrt{\Pi}\alpha^{2}+1\right)\right]$, \\
$A2= \Bigg[\alpha^{2}\Bigg(\sin^{2}\left(\Phi\right)/2-\csc^{2}\left(\Phi\right)/2  -\log\left(\sin^{2}\left(\Phi\right)\right)\Bigg)+\frac{B}{A}\Bigg]$\\
 $A3=\frac{2C}{\sqrt{(1-K)(K+Cr^{2})}} \alpha^{3}\left(\Pi\right)^{1/4}\left(\Upsilon\right)^{1/4}$,
  \\ \\
 $A4= \Bigg[\frac{\sin2\left(\Phi\right)}{4}+\csc^{2}\left(\Phi\right)\times \cot\left(\Phi\right)  -\cot\left(\Phi\right)\Bigg], ~~~and~~~  A5=\left(\Upsilon\right)^{1/4}\left[\left(\Pi\right)\alpha^{2}+\left(\Pi\right)^{1/2}\right], $ \\
and $\Pi =\frac{K+Cr^{2}}{1-K} ,~~~~\Upsilon= \frac{1+Cr^{2}}{1-K}$~~~ and
~~~$\Phi= \left(\arctan\left(\frac{K+Cr^{2}}{1-K}\right)^{1/4}\alpha\right). $

\begin{figure}[hbt!]
\centering
\includegraphics[width=6.5cm]{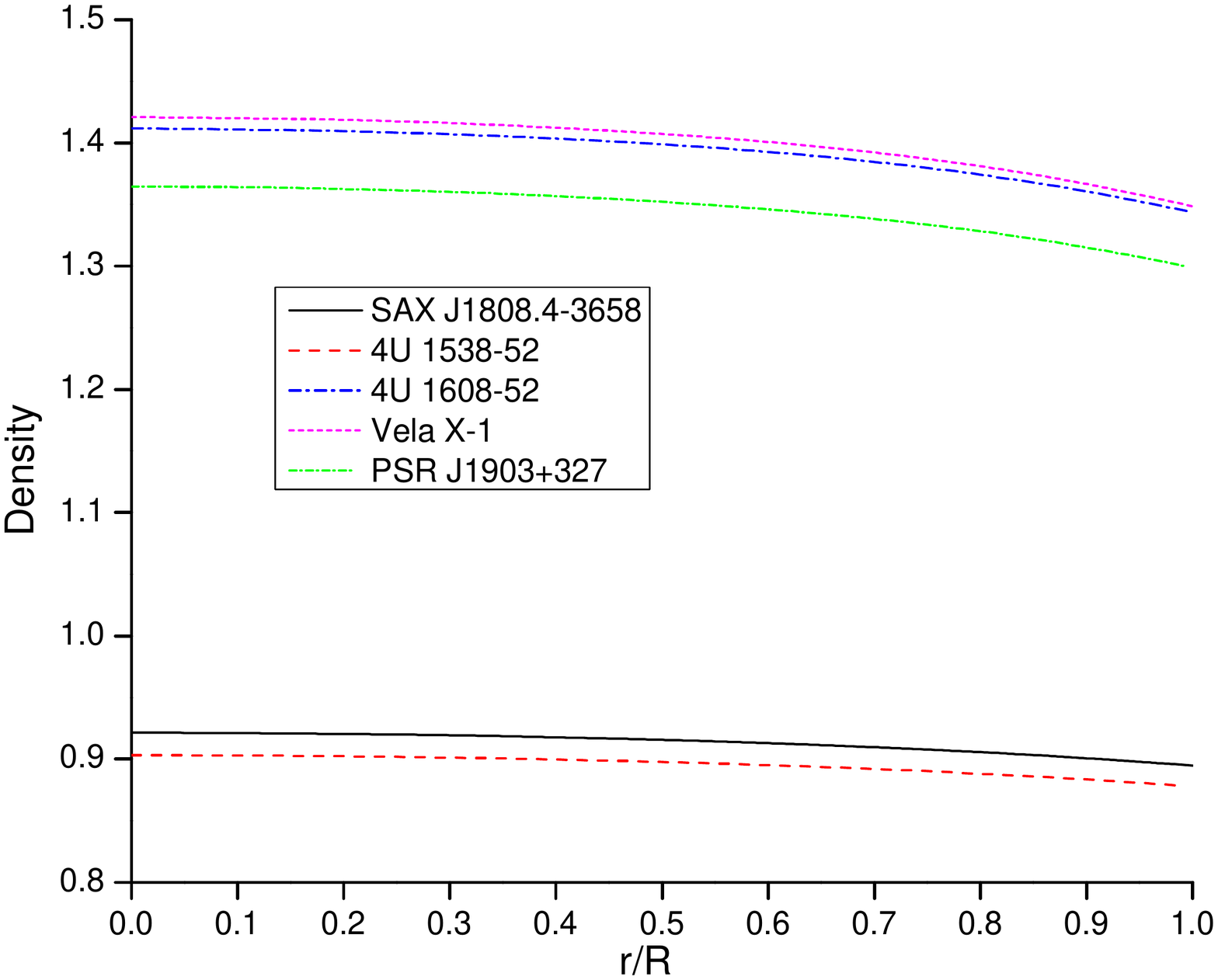} \includegraphics[width=6.5cm]{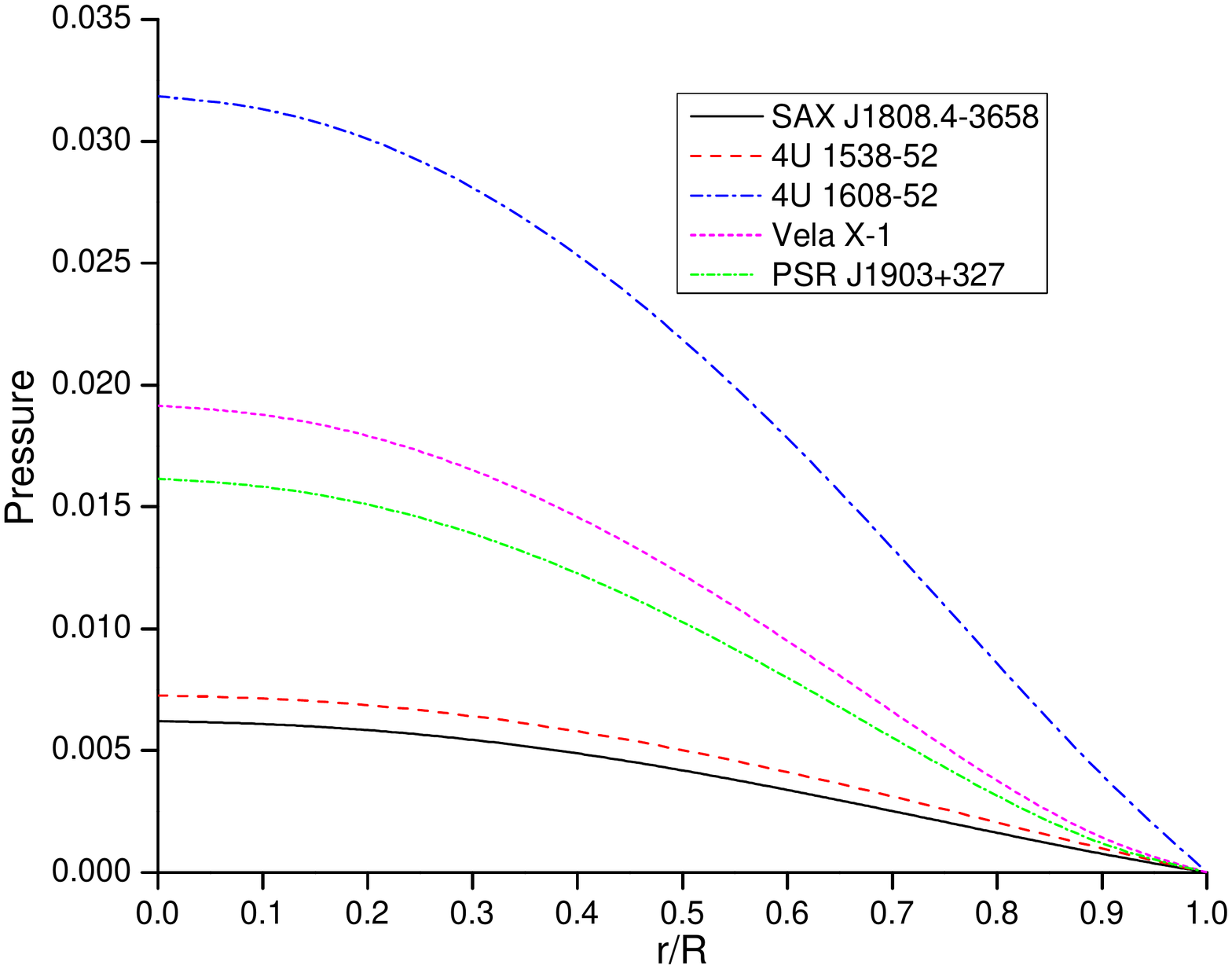}
\includegraphics[width=6.5cm]{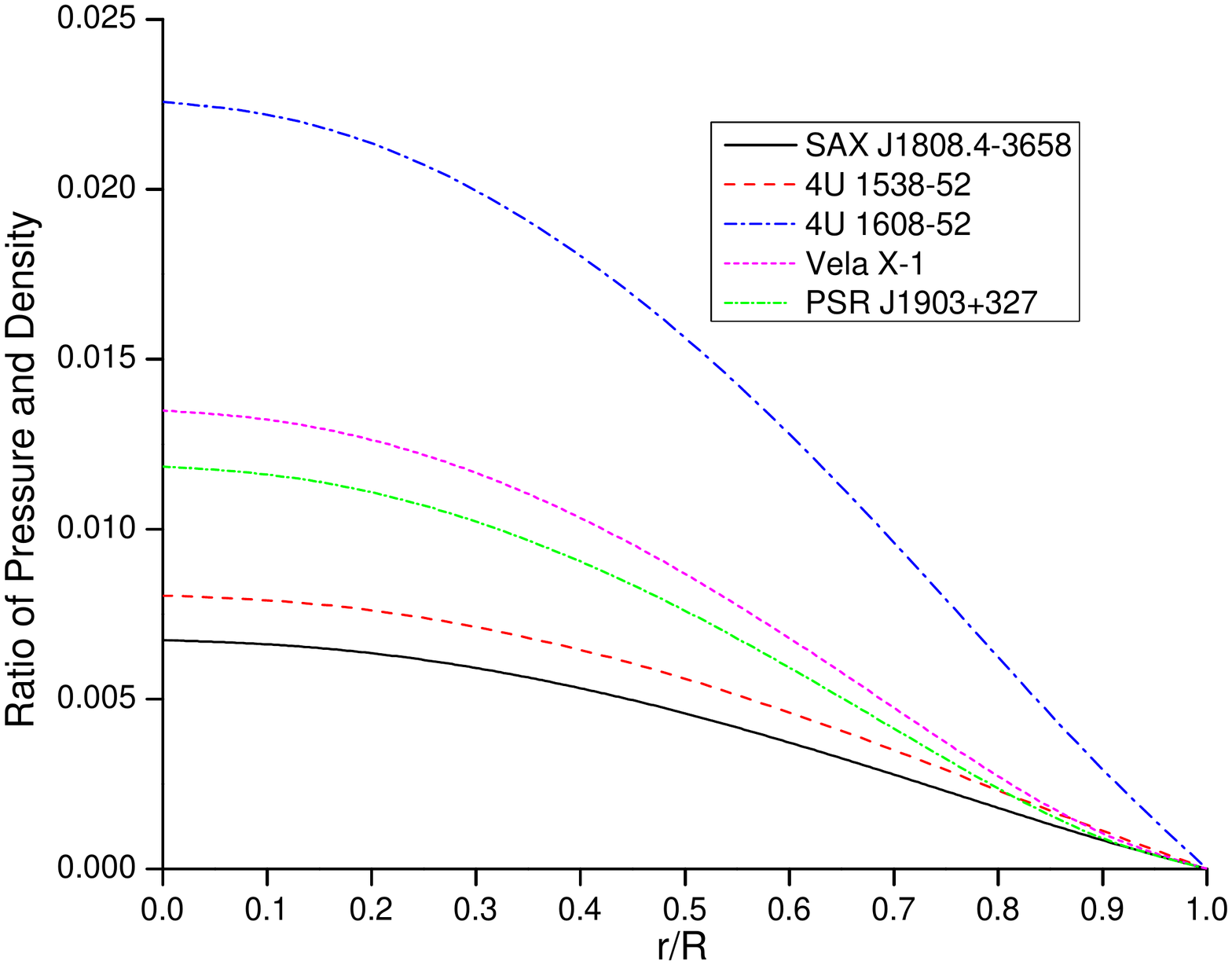} \includegraphics[width=6.5cm]{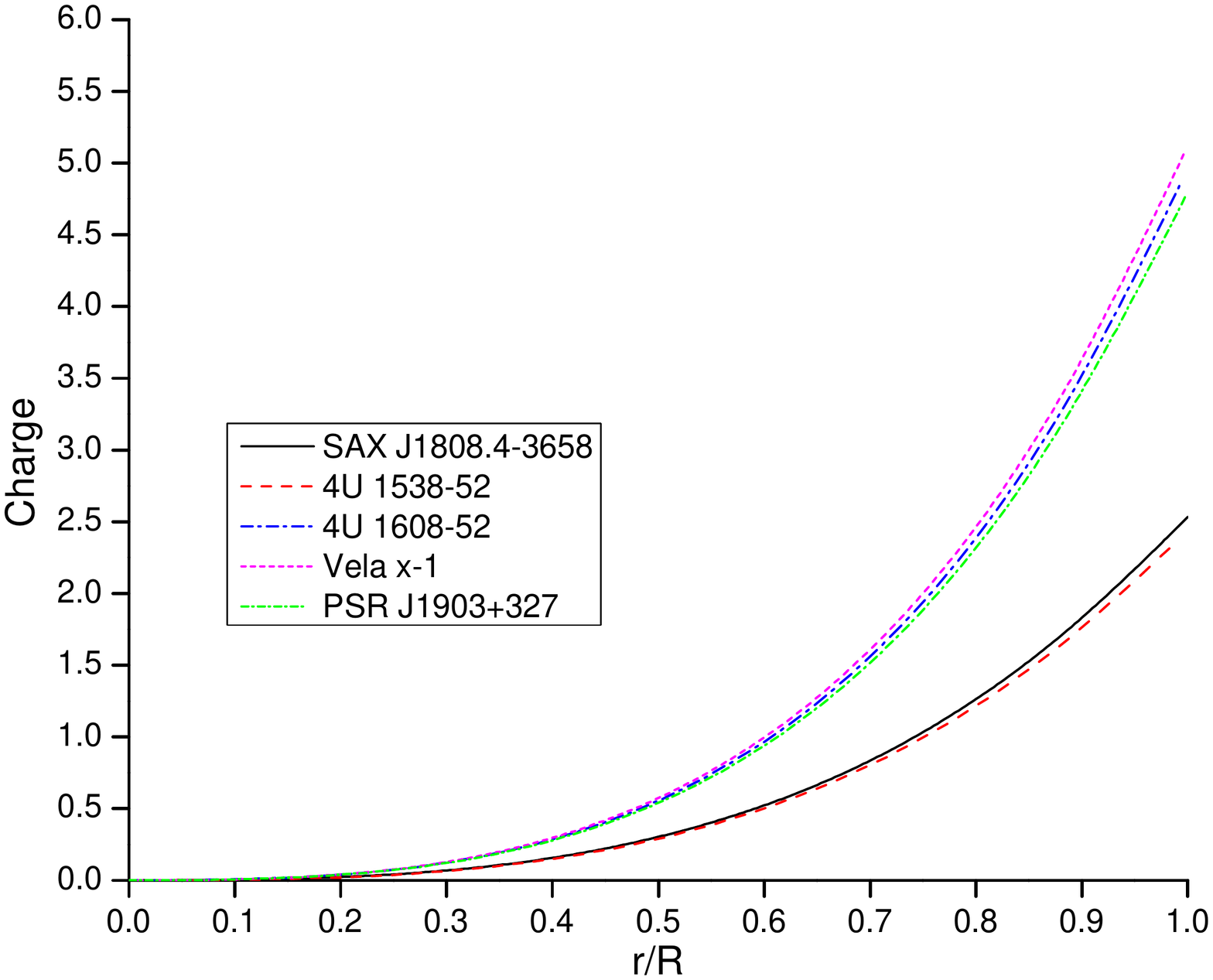}
\caption{\emph{Radial variation of energy density, pressure, density-pressure ratio and charge in their normalized forms inside the star have been plotted from top to bottom, for the compact objects SAX J1808.4-3658, 4U 1538-52, 4U1608-52, Vela X-1 and PSR J1903+0327.  The parameter values which we used for graphical presentation are (i) $K$ = 0.0000293, $ CR^2 = -0.000009$, $ \alpha^{2} $=54.3, $M$ = $0.9 M_{\odot}$, and $R$ = 7.951 km for SAX J1808.4-3658 (see Table~ \ref{table2}), (ii) $K$ = 0.00003, $CR^2= - 0.00000901$, $\alpha^{2}$ = 55, $M$ =$0.87 M_{\odot}$ and $R$ = 7.866 km for 4U 1538-52 (see Table~ \ref{table3}), (iii) $K$=0.00017, $CR^2 = - 0.00008$, $\alpha^{2}$=20.527, $M$ = $1.74M_{\odot}$ and $R$ = 9.3 km for 4U1608-52 (see Table~ \ref{table4}), (iv) $K$ = 0.00019, $CR^2 = - 0.00009$, $\alpha^{2}$=20.506, $M$ = $1.77M_{\odot}$ and $R$ = 9.56 for Vela X-1 (see Table~ \ref{table5}), and (v) $K$ = 0.0002, $CR^2=-0.000091$, $\alpha^{2}$= 20.18, $M$ = $1.667M_{\odot}$ and $R$ = 9.483 for PSR J1903+0327 (see Table~ \ref{table6}).}}
\label{f1}
\end{figure}

\begin{table}
\centering \caption{{ The surface density, mass, radius, ratio of mass-radius and value of $C$ for compact star candidate.} }
\scalebox{0.8}{
\begin{tabular}{c|c|c|c|c|c} \hline
 Compact Stars  & $R(Km)$ & $M(M_{\bigodot}$)    & Surface density ($g/cm^{3}$) & $-\,C$ (km $^{-2} $)  &  $M/R$ \\ \hline
  4U 1538-52 ~(Rawls et al. \cite{R}) & 7.866 & 0.87 & 7.586$ \times 10^{14} $ & 1.145$ \times 10^{-7} $ & 0.16314 \\ \hline
 4U 1608-52~ (Gu$\dot{v}$er et al. \cite{Guver}) & 9.3 & 1.74 & 8.333$ \times 10^{14} $ & 9.25$ \times 10^{-7} $ & 0.2694 \\ \hline
 Vela X-1~ (Rawls \textit{et al.} \cite{R}) & 9.56 & 1.77 & 7.914$ \times 10^{14} $ & 9.84$ \times 10^{-7} $ & 0.27309\\ \hline
 PSR J1903+327 ~ (Freire \textit{et al.} \cite{F}) & 9.438 & 1.667 & 7.821$ \times 10^{14} $ & 1.021$ \times 10^{-7} $ & 0.2605\\ \hline
 SAX J1808.4-3658 ~ (Elebert \textit{et al.}  \cite{E}) & 7.951 & 0.9 & 7.58$ \times 10^{14} $ & 1.421$ \times 10^{-7} $ & 0.16696 \\ \hline
 \label{table1}
\end{tabular}}
\end{table}

To obtain the models for a spherically charged star, in this study, the interior solution goes up to a certain radius $R$, say. To have a useful stellar model, it should satisfy the following conditions throughout the stellar configuration:
\begin{itemize}
\item The spacetime is assumed not to possess an event horizon.
\item To keep the centre of the spacetime regular, energy density $\rho$ and pressure $p$ are positive within the radius.
\item $ (d\rho/dr)_{r=0}=0 $ and $ (d^{2}\rho/dr^{2})_{r=0}<0, $ so that density gradient  $ d\rho/dr $ is negative within $ 0<r < R $.
\item $ (dp/dr)_{r=0}=0 $ and $ (d^{2}p/dr^{2})_{r=0}< 0, $ so that pressure gradient $ dp/dr $ is negative within $ 0< r < R $.
\end{itemize}

Since analyzing the structure of charged spheres made of perfect fluids, we plot the results in terms of various values of the arbitrary constant $K$ and $C$. We plot the density, pressure, density-pressure ratio and the total charge $Q$  of the stars at the surface as a function of their radius. Fig. (1) (in the manuscript), therefore, conveys an important message: at the interior solution, the density and pressure are maximum at the center and monotonically decreasing towards the boundary. It is the Buchdahl criterion \cite{Buchdahl} for a stable stellar structure.  This argument, even if formulated in the isotropic and anisotropy one can find that density and pressure are increasing towards the boundary if $C<0$ while it is negative if $C>0$ for $0<K<1$. However, In the case of the modified theory of gravity, in particular, $f(R,T)$ gravity, the density profile is also monotonically increasing for some range of the matter coupling parameters. It means that charged solutions of Einstein-Maxwell field equations lead to the determining more stable stellar structure where Coulomb repulsion works against gravitational attraction. The key aspect of the problem is that this given methodology gives a physically valid solution only for the charged case when $0< K<1$ and $C< 0$.

In addition to this we verify that pressure vanishes at the boundary, as a function of the radial coordinate $r$ (see \ref{f1}). Here it is observed that electric field associated with Eq. (\ref{14}) vanishes at $r = 0$. We plot the radius of the resulting spheres as a function of the charge distribution in Fig. (\ref{f1}) for different compact stars. This means vanishing of the electric field at the center of a spherically symmetric charge distribution remains regular and positive throughout the sphere Table \ref{table1}.

\section{Boundary conditions} \label{bou}
The expression (\ref{18}),  (\ref{19}) and  (\ref{14}) represents the interior solution up to a certain radius until the null pressure point, at its surface, where we match with the Reissner-Nordstr$\dot{{\text{o}}}$m external solution. At this stage one puts forward some sensible requirements e.g., based on data for different radii and masses of compact stars that are used to fix the values of constants $\alpha$, $C$ and $K$. The exterior vacuum solution is then given by
\begin{eqnarray}
 ds^{2}= \bigg( 1-\frac{2M}{r}+\frac{q^{2}}{r^{2}}\bigg)dt^{2}
 -\bigg(1-\frac{2M}{r}+\frac{q^{2}}{r^{2}}\bigg)^{-1}dr^{2}-r^{2}(d\theta^{2}+\sin^{2}\theta d\phi^{2}),~~\label{20}
\end{eqnarray}
where $ M $ and $Q$ are the total gravitational mass and charge of the fluid sphere. For this purpose, the total mass can be written as
\begin{eqnarray}
M=\zeta(R)+\xi(R), \label{21}
\end{eqnarray}
with the definition $\zeta(R) = \frac{\kappa}{2}\int_{0}^{R}\rho r^{2}dr$, $\xi(R)= \frac{\kappa}{2} \int_{0}^{R}r\sigma  qe^{\lambda/2}dr$ and $Q = q(R)$ represents, the mass within the sphere, the mass equivalence of the electromagnetic energy of distribution and $Q$ is the total charge inside the fluid sphere (see also \cite{Florides} for a review).  We shall model compact stars by matching the interior solution, governed by the Eq. (\ref{1}) to an exterior Reissner-Nordstr$\dot{{\text{o}}}$m vacuum solution (\ref{20}), at the boundary surface $r = R$. The imposition of smooth boundary conditions on the boundary surface with the following relations
 \begin{eqnarray}
 e^{-\lambda}=1-\frac{2M}{R}+\frac{Q^{2}}{R^{2}}, ~~~ and~~~
 e^{\nu}= y^{2}=1-\frac{2M}{R}+\frac{Q^{2}}{R^{2}},\label{22}
 \\p(R)=0,\label{27}~~~ and~~~ q(R)=Q.\label{23}
\end{eqnarray}
Note that across the boundary $r = R$ of the star together with the condition that pressure vanish at the surface $p(R)$ = 0 help us to determine these constants.  From hence several parameters of the model are going to fix by using the full set of boundary conditions. In the following analysis the values of constant coefficients $\alpha$, $C$ and $K$ are determined for some particular values of observed 
massive stellar objects such as SAX J1808.4-3658, 4U 1538-52, 4U 1608-52, Vela X-1 and PSR J1903+327, respectively.

With the purpose of determining surface gravitational redshift $z_S$  for different compact objects by using the formal definition $z_S$ = $\Delta \lambda/\lambda_{e}$ = $\frac{\lambda_{0}-\lambda_{e}}{\lambda_{e}}$, where $\lambda_{e}$ is the emitted wavelength at the surface of a nonrotating star and $\lambda_{0}$ is the observed wavelength received at radial coordinate $r$.
The gravitational redshift $z_S$ within a static line element is given by
\begin{eqnarray}
\label{eq24}
1+z_S = \arrowvert g_{tt}(R) \arrowvert ^{-1/2} = \left(1-\frac{2M}{R}+\frac{Q^{2}}{R^{2}}\right)^{-1/2},
\end{eqnarray}
where $g_{tt}(r)$ = $e^{\nu(R)}$ =$\left(1-\frac{2M}{R}+\frac{Q^{2}}{R^{2}}\right)$ is the metric function.
It has been shown \cite{Buchdahl,Straumann}
that for spherically symmetric perfect fluid spheres the gravitational redshift is $z_{s} <$ 2, and for anisotropic case this values turns out to be 3.84, as in \cite{Karmakar,Barraco}. But, Boehmer and Harko \cite{Boehmer} showed that this value could be increased
up to $z_{s} \leq$ 5, which is consistent with the bound  $z_{s} \leq$ 5.211 obtained by Ivanov \cite{Ivanov}. But for quasiblack hole the redshift at the surface of the star is indefinitely large i.e. the numerical value of the order of 100 (see reference \cite{arbanil}). In each Tables~ \ref{table2}-\ref{table6}, we enlisted the values of  redshift for different compact objects by taking the same values, which we have used for graphical presentation in Fig.\,\ref{f1}. For each compact object we found that the value of surface redshift is maximum at the centre and monotonic decreasing towards the boundary. In the left panel of Fig.\,\ref{f2}, we provide graphs of surface redshift ($z_S$) for the strange star candidates SAX J1808.4-3658, 4U 1538-52, 4U 1608-52, Vela X-1 and PSR J1903+327, which yield fall $z_{s} <$ 2 in every cases.

\section{Structure properties of compact objects}
\label{phy}

In this section we introspect more details about the stellar configuration by performing  some analytical calculations and studying  physical properties of the interior of the fluid sphere. 
We are assuming here how the TOV equations are modified by the presence of electric field and contributes to the dynamics of the self-gravitating system. Finally, we develop our considerations by investigating the type of compact objects that might arise from these solutions and to restrict the model arbitrariness.

\subsection{Causality condition}
To confirm that we are not losing essential physics i.e. the speed of sound propagation $v^2_s = dp/ d\rho$, at the interior of the star. From a mathematical point of view, the velocity of sound is less than the velocity of light. In essence of this we fix
$c = 1$, and investigate the speed of sound for charged fluid matter. For the
stability of our model, we shall adopt Herrera's overtuning technique \cite{Herrera(2016)}, which states that the region of the interval should be $0 < v^{2}=\frac{dp}{d\rho} < 1$. Now employing the
Eqs. (\ref{18}) and Eq. (\ref{19}), we get
 \begin{eqnarray}
 \frac{dp}{c^{2}d\rho}=\frac{\frac{N(r)}{(N_{6}(r).A5)^{2}}-\frac{2C^{2}r(1-K)}{K(1+Cr^{2})^{2}}+N_{2}(r)+N_{3}(r)}{\bigg(N_{1}(r)-N_{2}(r)-N_{3}(r)\bigg)},\label{25}
 \end{eqnarray}

 \begin{figure}[hbt!]
\centering
\includegraphics[width=7cm]{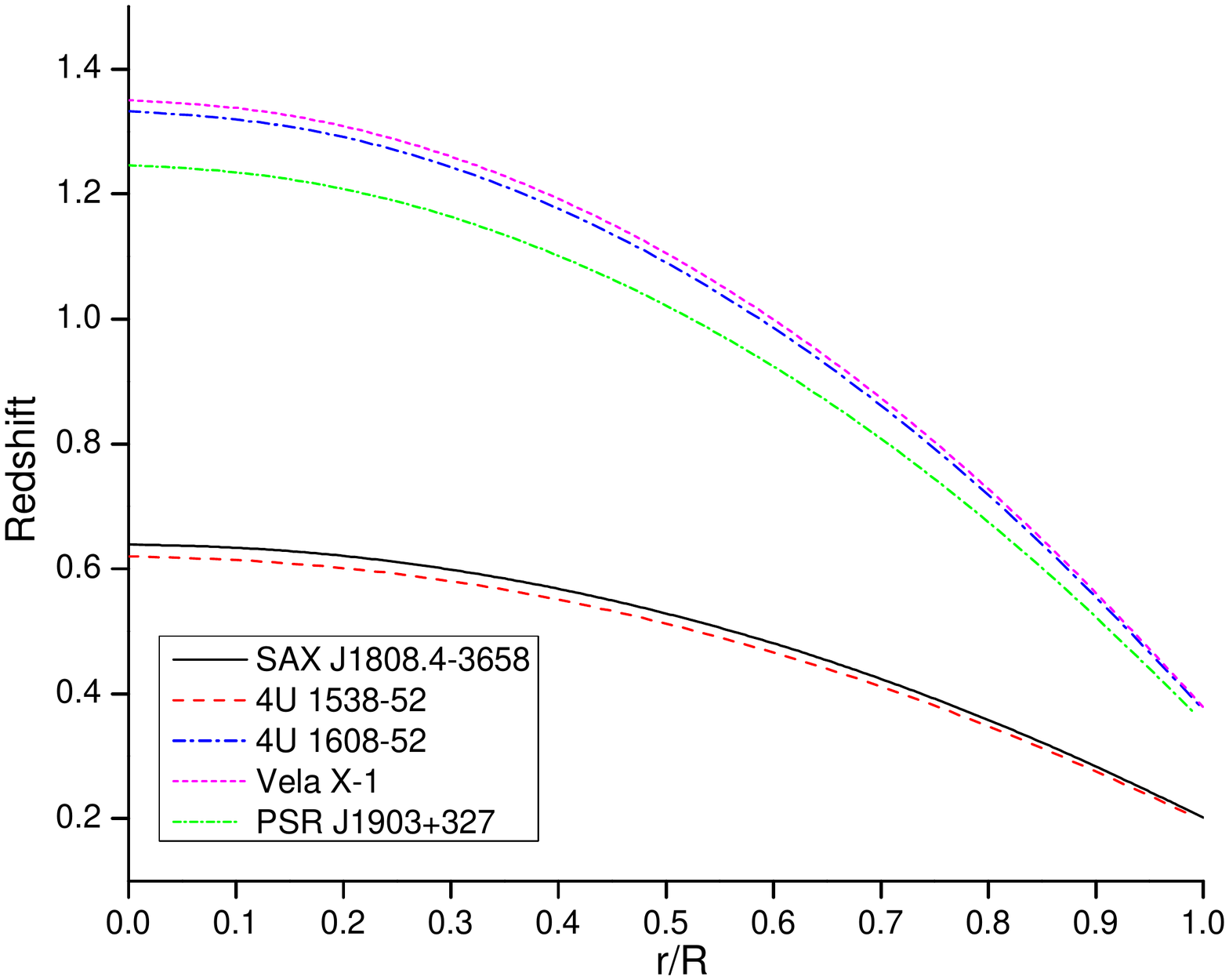} \includegraphics[width=7cm]{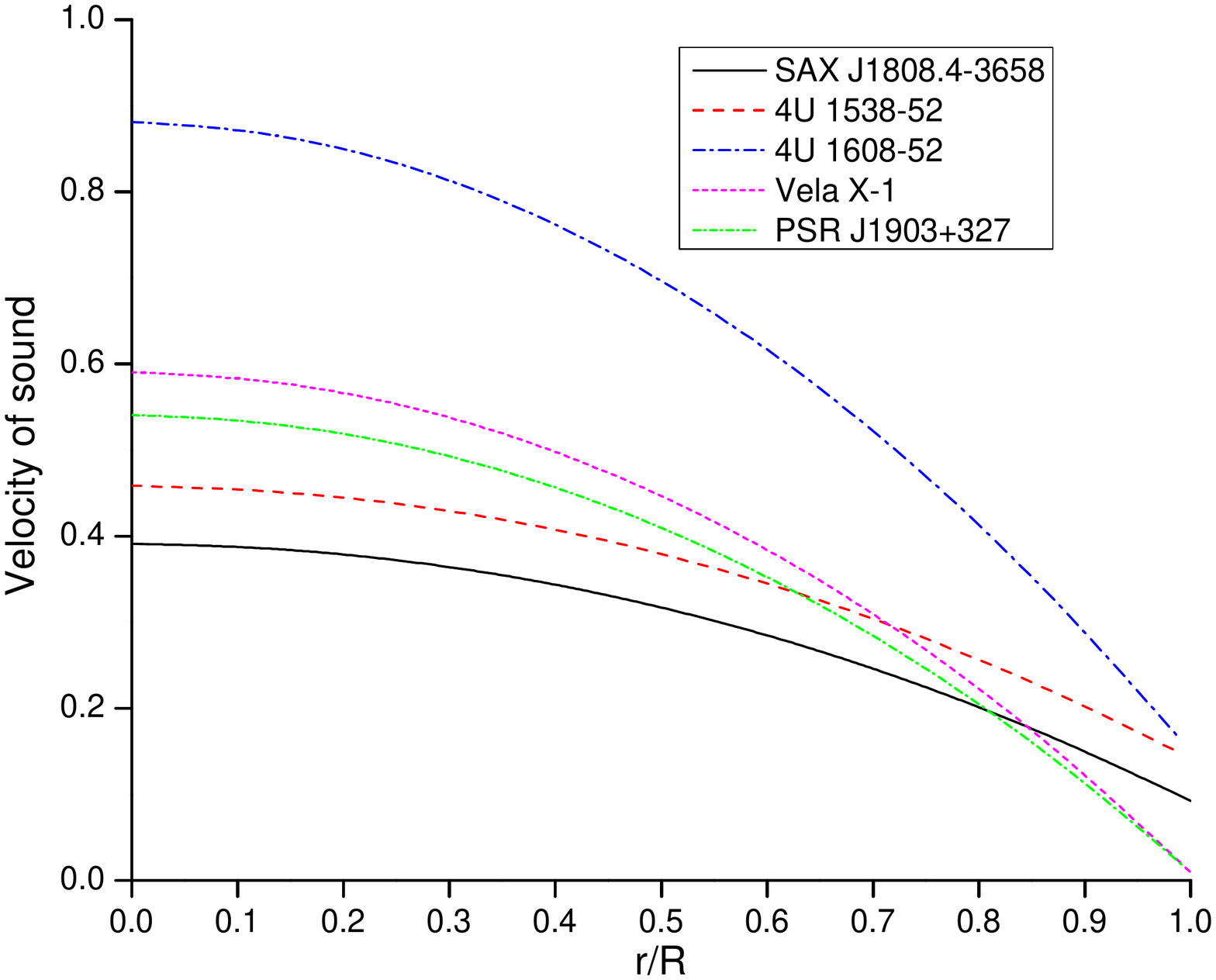}
\caption{\emph{Behaviour of redshift and sound velocity have been plotted
as a function of the radial coordinate $r/R$ for different compact star candidates. In a given plot, we use same data as of Fig.\,\ref{f1}}}
\label{f2}
\end{figure}
where the notations are\\
  $ N(r)=A2. A5\Bigg[N_{5}(r)\bigg( N_{6}(r). A2+A13. A4\bigg)+N_{4}\Bigg(\bigg(N_{7}(r)+N_{8}(r)\bigg)A2+N_{6}(r).N_{9}(r)+A13.N_{10}(r)+A4.N_{11}(r)\Bigg)\Bigg]-N_{4}(r)\bigg( N_{6}(r).A2+A13.A4\bigg)\bigg(A5.N_{9}(r)+A2.N_{12}(r) \bigg)$,\\
$ N_{1}(r)=\frac{2C^{2}r(1-K)(5+Cr^{2})}{K(1+Cr^{2})^{3}}$,~~~~~$A13=\alpha^{3}\left(\Pi\right)^{1/4}\left(\Upsilon\right)^{1/4}$\\ \\
 $\resizebox{1\hsize}{!}{$ N_{2}(r)=\frac{C^{2}r(\psi(1-Cr^{2})+rt(r)(1+Cr^{2}))}{8K(1+Cr^{2})^{3}(1-K)}\times\bigg[4(1+Cr^{2})^{2}+(1-K)(2+7Cr^{2}+K(1-4 Cr^{2}))\bigg],~~$}$ \\ \\
 $\resizebox{1\hsize}{!}{$t(r)=\frac{2\alpha^{2}\Big[\frac{2Cr\alpha^{2}}{1-K}+\Big(\Pi\Big)^{-1/2}\frac{Cr}{1-K}\Big]}
 {\Big[\Pi\alpha^{2}+ \sqrt{\Pi}\Big]^{2}},~~
 N_{3}(r)=\frac{C^{2}r^{2}\psi}{8K(1+Cr^{2})^{2}} \times\bigg[ 16Cr \Upsilon+(14Cr-8KCr)\bigg],~~$}$ \\ \\
 $N_{4}(r)=\frac{2C(K+Cr^{2})}{K(1+Cr^{2})\big( (1-K)(K+Cr^{2})\big)^{1/2}},~~~~~~~~
 N_{5}(r)=\frac{2C^{2}r(1-2K-Cr^{2})}{K(1-K)^{1/2}(1+Cr^{2})^{2}(K+Cr^{2})},\\ \\
 N_{6}(r)=\left[\frac{\left(\Pi\right)^{3/2}\alpha^{2}+\Pi}{2\left(\Upsilon\right)^{3/4}}+\left(\Upsilon\right)^{1/4}\left(2\sqrt{\Pi}\alpha^{2}+1\right)\right] \\ \\
 N_{7}(r)=\frac{Cr(1-K)^{3/4}}{(1+Cr^{2})^{7/4}}\Bigg[ \Upsilon\bigg(\frac{3}{2}\sqrt{\Pi}\alpha^{2}+1\bigg)-\frac{3}{4}\Pi\bigg(\sqrt{\Pi}\alpha^{2}+1\bigg)\Bigg],\\ \\
 N_{8}(r)=\frac{2Cr}{1-K}\bigg(\frac{1}{\Upsilon}\bigg)^{3/4}\Bigg[(\Pi)^{-1/2}\alpha^{2}\Bigg( \frac{1}{2}(\Pi)+\Upsilon \Bigg)+\frac{1}{4}\Bigg],~~~~~~
 N_{9}(r)=\frac{\alpha^{3}Cr}{1-K}\left(\frac{A4}{(\Pi)^{3/4}+(\Pi)^{5/4}\alpha^{2}}\right),$\\ \\
$\resizebox{1\hsize}{!}{$ N_{10}(r)=\frac{1}{2(1-K)}\left(\frac{\alpha Cr}{(\Pi)^{3/4}+(\Pi)^{5/4}\alpha^{2}}\right)\Bigg[\cos2\left(\Phi\right)/2-2\csc^{2}\left(\Phi\right)\times \cot^{2}\left(\Phi\right)-2\csc^{4}\left(\Phi\right)+\csc^{2}\left(\Phi\right)\Bigg],~~$}\\ \\
N_{11}(r)=\frac{\alpha^{3}Cr}{2(1-K)}\left[ (\Upsilon)^{-3/4}(\Pi)^{1/4}+(\Upsilon)^{1/4}(\Pi)^{-3/4}\right],\\ \\
N_{12}(r)=\frac{Cr}{2(1-K)}(\Upsilon)^{-3/4}(\Pi)^{1/2}\left( (\Pi)^{1/2}\alpha^{2}+1\right)+(\Upsilon)^{1/4}\frac{2Cr}{1-K}\times\Bigg(\alpha^{2}+\frac{(\Pi)^{-1/2}}{2} \Bigg). $

 Here, we have used the same notation for $\Phi$ and $\Upsilon$, as before. In order to carry out analytic calculations and to obtain
a better understanding of numerical results, we consider simple graphical representation in the right panel of Fig.\,\ref{f2}. One can also verify this by inspecting Eq. (\ref{25}). Thus, by carefully observing the  Fig.\,\ref{f2}, one arrives at the following conclusion that  velocity of sound lies within the proposed range for different compact objects. Interestingly, it is also evident that the slope for $\frac{dp}{d\rho}$
with charge is decreases towards the boundary.  For the
purposes of this calculation, we use the same parametric values as given in Fig.\,\ref{f1}.

\subsection{ Tolman-Oppenheimer-Volkoff (TOV) equations}
It is important to analysis the equilibrium conditions using
the Tolman-Oppenheimer-Volkoff (TOV) equation. The TOV equation mainly
constrains the structure of a static spherically symmetric body
which is in static gravitational equilibrium. Here we address how gravitational and other  fluid forces counteract with increasing electrostatic repulsion towards the boundary, where the pressure gradients tend to vanish. Now, by employing the generalized-TOV equation \cite{Tolman,Volkoff} in the presence of charge, as prescribed in \cite{Ponce}, we have the following form

\begin{figure}[hbt!]
\centering
\includegraphics[width=6.5cm]{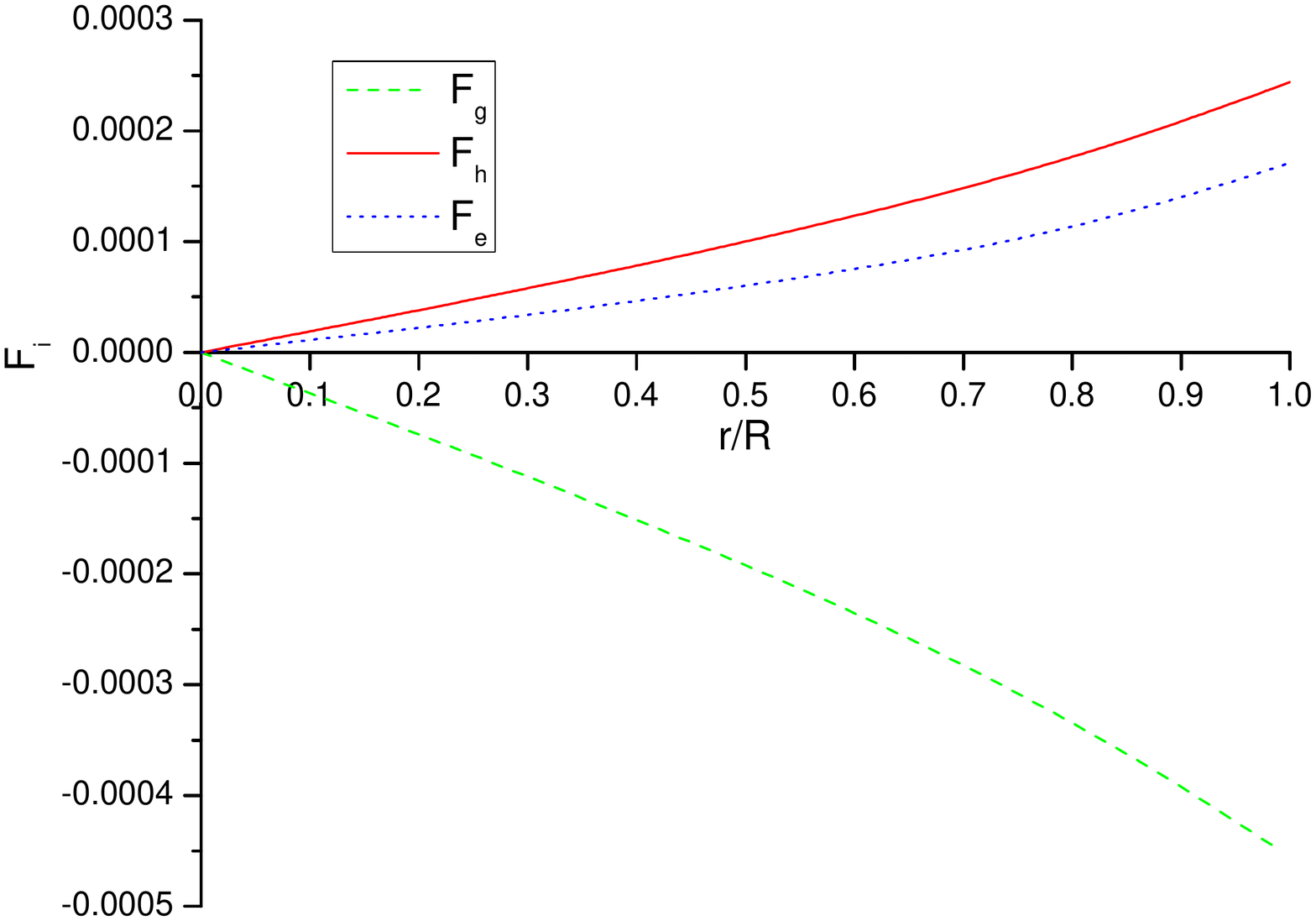} \includegraphics[width=6.5cm]{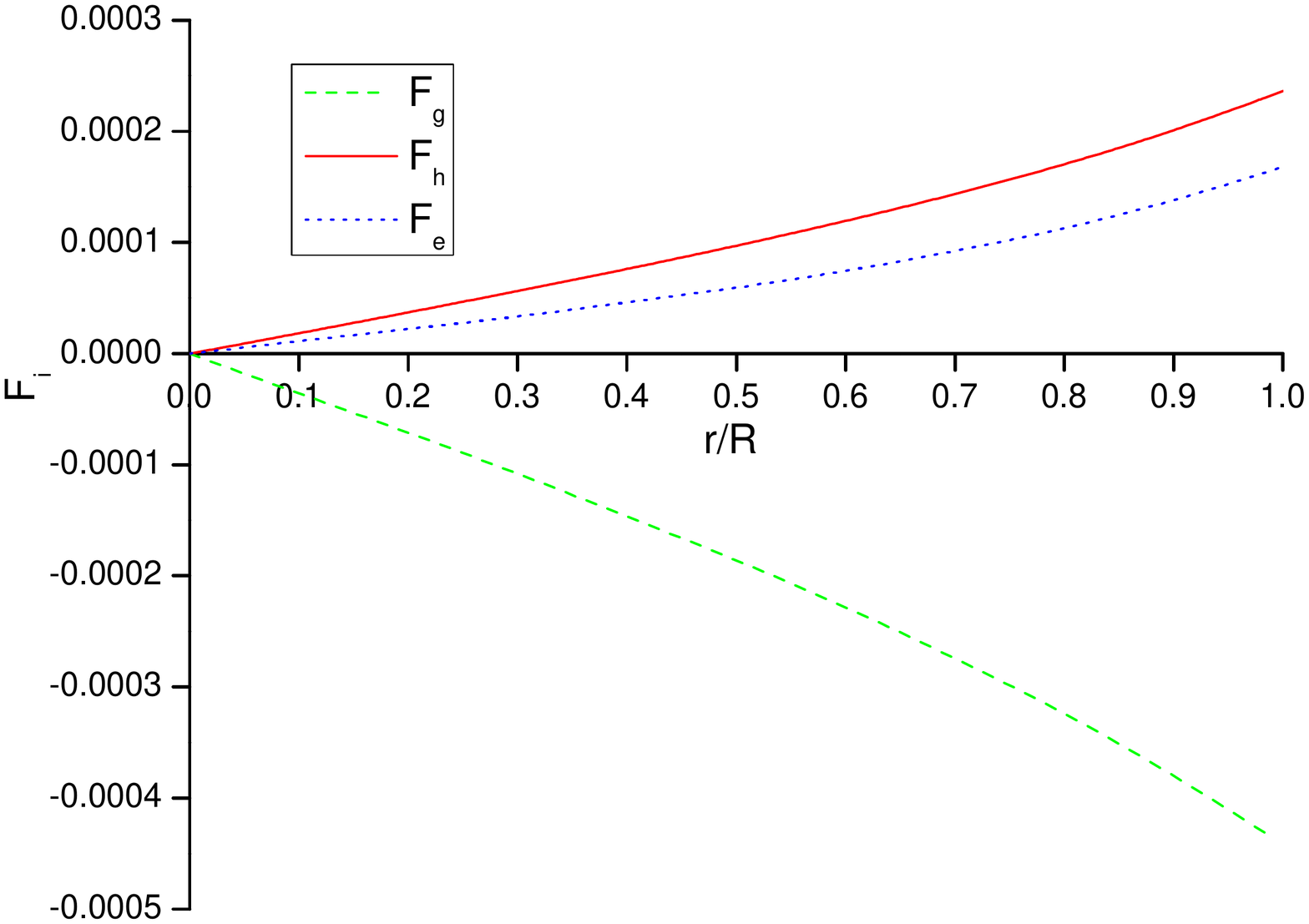}
 \includegraphics[width=6.5cm]{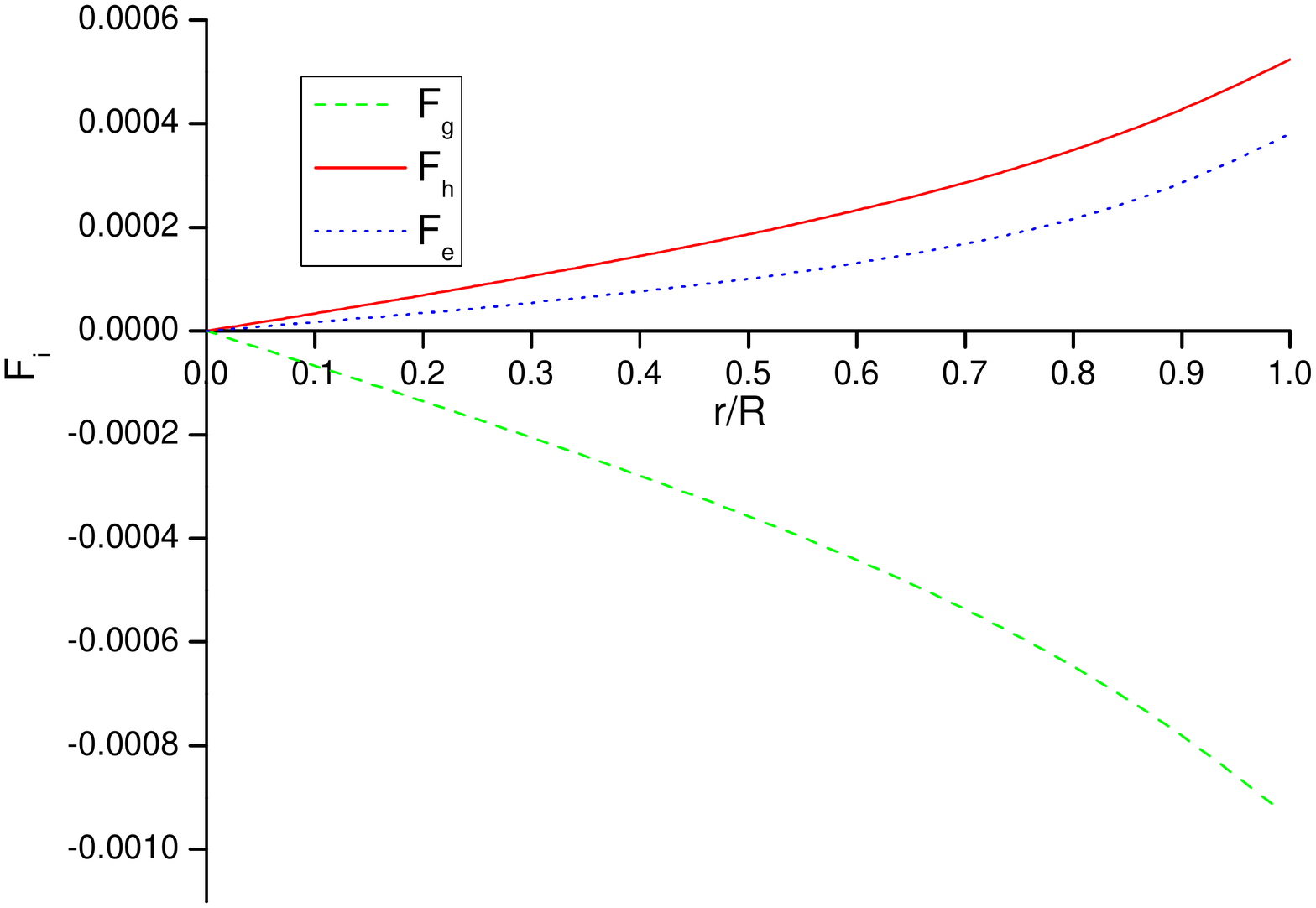} \includegraphics[width=6.5cm]{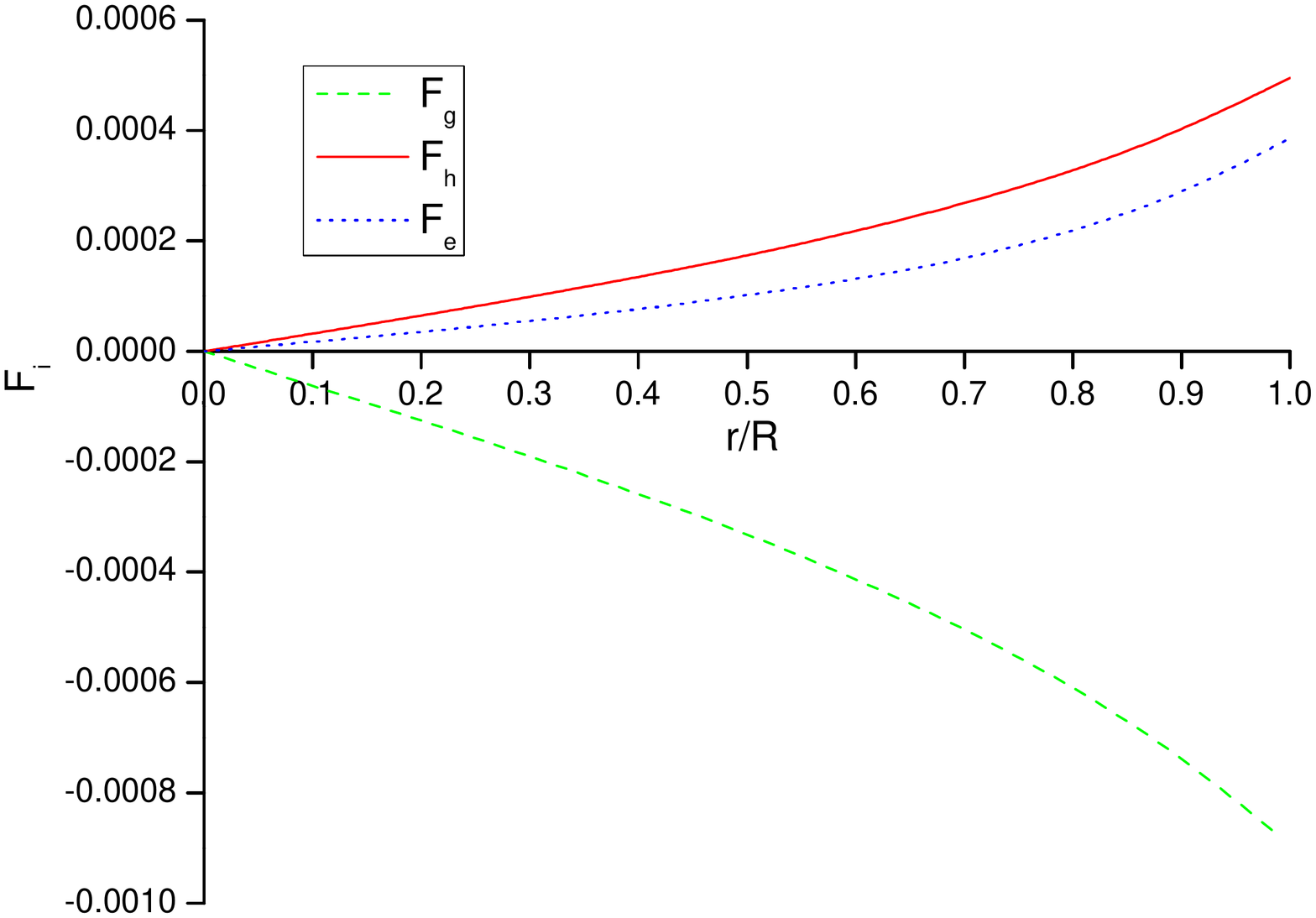}
\includegraphics[width=6.5cm]{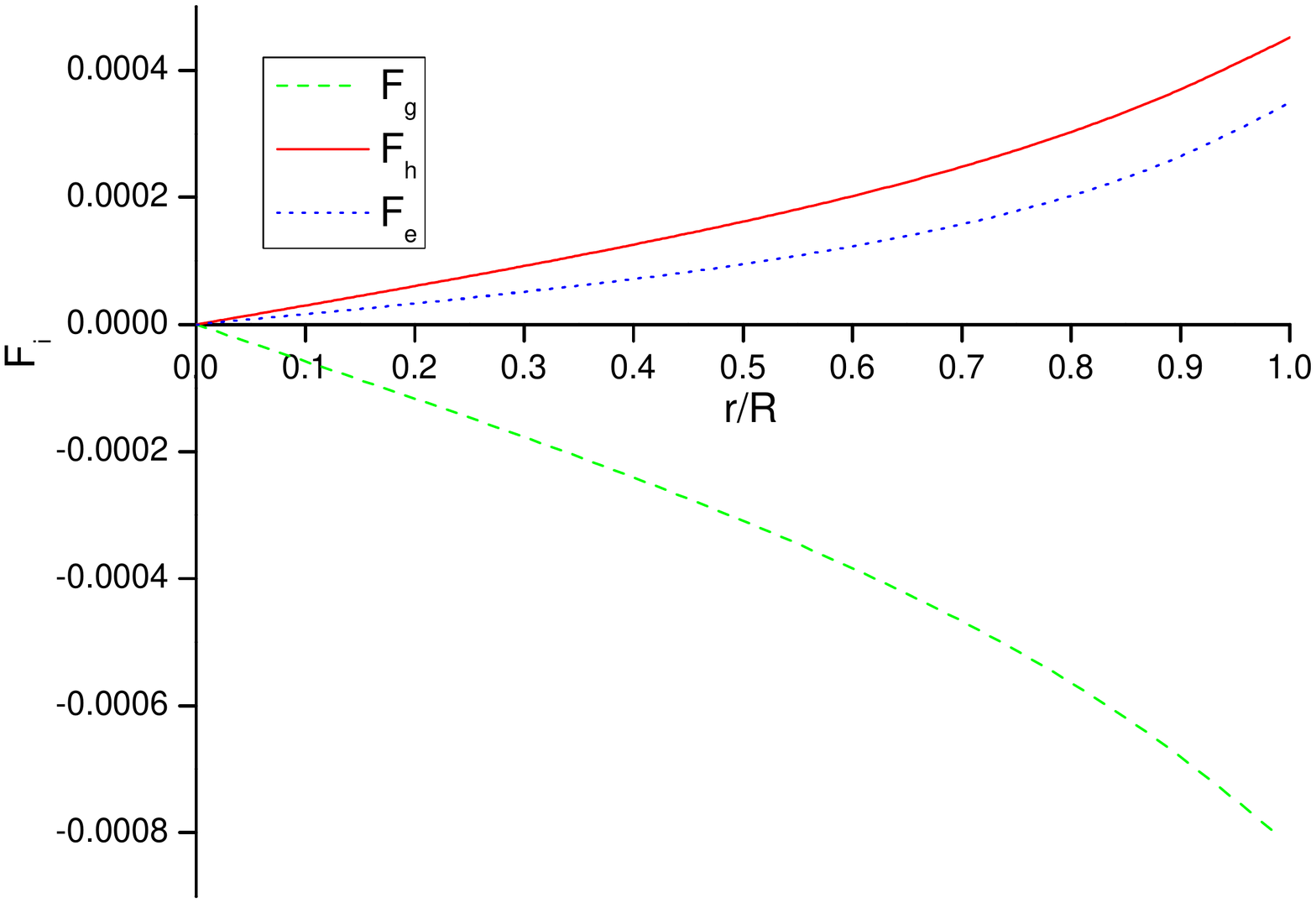} 
\caption{\emph{The plot depicts the different forces, namely, hydrostatic $(F_h)$, electric $(F_e)$ and gravitational $(F_g)$ forces, respectively.  Here we study
the effect of forces to maintain the stability of compact stars. See the text for details.}}
\label{f3}
\end{figure}

\begin{eqnarray}
 -\frac{M_G(\rho+p)}{r^2}e^{\frac{\lambda-\nu}{2}}-\frac{dp}{dr}+
 \sigma \frac{q}{r^2}e^{\frac{\lambda}{2}} =0,\label{26}
 \end{eqnarray}
where $M_G = M_G(r)$ is the effective gravitational mass within the radius r,
and q = q(r) is given by (\ref{14}). The expression for the effective gravitational mass is given by
\begin{eqnarray}
M_G(r)=\frac{1}{2}r^2 \nu^{\prime}e^{(\nu - \lambda)/2}.\label{27a}
\end{eqnarray}
Now, plugging the value of $M_G(r)$ in Eq. (\ref{25}), we get
\begin{eqnarray}
-\frac{\nu'}{2}(\rho+p)-\frac{dp}{dr}+\sigma \frac{q}{r^2}e^{\frac{\lambda}{2}} =0,  \label{28}
\end{eqnarray}
Let us now attempt to explain the above equation from an equilibrium point of view,
where the expression treated as a combination of three different forces namely gravitational $(F_g)$, hydrostatic $(F_h)$ and electric forces $(F_e)$, respectively. For our system the forces are as follows:
\begin{eqnarray}
F_g =-\frac{\nu'}{2}(\rho+p)=\frac{Z'}{2\pi Z}(\rho+p),\label{29}
\end{eqnarray}
\begin{eqnarray}
F_h=-\frac{dp}{dr}=-\frac{1}{8\pi}\left[\frac{N(r)}{(N_{6}(r).A5)^{2}}-\frac{2C^{2}r(1-K)}{K(1+Cr^{2})^{2}}+N_{2}(r)+N_{3}(r)\right],  \label{30}
\end{eqnarray}
\begin{eqnarray}
F_e=\sigma \frac{q}{r^2}e^{\frac{\lambda}{2}}= \frac{1}{8\,\pi\,r^4}\,\frac{dq^2}{dr}=\frac{1}{4\pi}\left[N_{2}(r)+N_{3}(3)\right],\label{31}
\end{eqnarray}
where we use the same notation as mentioned above.
The generalized TOV equations of a charged spherical body is shown in Fig.\,\ref{f3}. It may be mentioned here that static equilibrium configurations do exist due to the combined effect of hydrostatic force $(F_h)$ and electric force $(F_e)$, which is counterbalanced by the  gravitational force $(F_g)$. In order to check the condition, we have plotted Fig.\,\ref{f3},  for the set compact star candidates SAX J1808.4-3658, 4U 1538-52, PSR J1903+327, Vela X-1 and 4U1608-52, for the same values as mentioned  in Fig.\,\ref{f1}.

\begin{table}
\centering \caption{Structural properties of  strange star "SAX J1808.4-3658" within radius}
\scalebox{0.8}{%
\begin{tabular}{c|c|c|c|c|c|c}
\hline \textbf{Radius} & \multicolumn{6}{c}{Values of the physical parameters$ \quad K=0.0000293, \quad CR^2 = -0.000009, \quad \alpha^{2} = 54.3$}   \\ \hline
x=$\frac{r}{R}$ & Pressure (P) & Density (D) & 	Charge  ($q_1$)  &  $dp/c^{2}d\rho$ &	 P/D  & Redshift ($Z_S$)  \\ \hline
0&	0.006203&	0.921475&		0&	0.391163&	0.006731&	0.639357\\ \hline
0.1&	0.006119&	0.92126&		0.00227&	0.388247&	0.006642&	0.634983\\ \hline
0.2&	0.00587&	0.920611&		0.018212&	0.379496&	0.006376&	0.621863\\ \hline
0.3&	0.005462&	0.919514&		0.061761&	0.364892&	0.00594&	0.599999\\ \hline
0.4&	0.004904&	0.91794&		0.1474&	0.344405&	0.005342&	0.569392\\ \hline
0.5&	0.004212&	0.915852&		0.290497&	0.317984&	0.004599&	0.530042\\ \hline
0.6&	0.00341&	0.913191&		0.507714&	0.285551&	0.003734&	0.481943\\ \hline
0.7&	0.002529&	0.909881&		0.817561&	0.246981&	0.002779&	0.425078\\ \hline
0.8&	0.001616&	0.905811&		1.241159&	0.202092&	0.001784&	0.359414\\ \hline
0.9&	0.000739&	0.900831&		1.803365&	0.150606&	0.00082&	0.284888\\ \hline
1&	0&	0.89472&		2.534498&	0.092109&	0& 0.20139
 \\ \hline
\label{table2}
\end{tabular}}
\end{table}

\begin{table}
\centering \caption{Structural properties of strange star "4U 1538-52" within radius }
\scalebox{0.8}{%
\begin{tabular}{c|c|c|c|c|c|c}
\hline \textbf{Radius} & \multicolumn{6}{c}{Values of the physical parameters $ K=0.00003, \quad CR^2=-0.00000901, \quad \alpha^{2}=55$}  \\ \hline
x=$\frac{r}{R}$ & Pressure (P) & Density (D) & 	Charge  ($q_1$)  &  $dp/c^{2}d\rho$ &	 P/D  & Redshift ($Z_S$)  \\ \hline
0&	0.004184&	0.900973&		0&	0.300104&	0.004644&	0.619595\\ \hline
0.1&	0.004122&	0.900764&		0.002216&	0.297574&	0.004576&	0.615365\\ \hline
0.2&	0.003936&	0.900132&		0.017776&	0.289977&	0.004373&	0.602678\\ \hline
0.3&	0.003633&	0.899063&		0.060278&	0.277293&	0.004041&	0.581535\\ \hline
0.4&	0.003222&	0.897531&		0.143842&	0.259482&	0.00359&	0.551938\\ \hline
0.5&	0.002718&	0.8955&		0.283434&	0.236485&	0.003035&	0.513887\\ \hline
0.6&	0.002144&	0.892914&		0.495251&	0.208209&	0.002401&	0.467377\\ \hline
0.7&	0.001529&	0.8897&		0.797248&	0.174521&	0.001718&	0.412393\\ \hline
0.8&	0.000919&	0.885755&		1.209848&	0.135223&	0.001037&	0.348904\\ \hline
0.9&	0.000377&	0.880936&		1.756989&	0.090032&	0.000428&	0.276851\\ \hline
1&	0&	0.87504&		2.467715&	0.038534&	0&	0.196133\\ \hline
\label{table3}
\end{tabular}}
\end{table}

\subsection{Energy conditions}

Given the fact that stellar structures are supported by perfect fluid distributions of matter,  we shall specify the energy conditions according to classical field theories of gravitation. One can assume that a relation which demands that matter density and pressure obeying certain restrictions. Note that the superluminal censorship theorem \cite{Olum,Bassett} positive mass theorem \cite{Schoen}, singularity theorems \cite{Hawking}, and various constraints on black hole surface gravity \cite{Visser} have profound and far-reaching import applications of some types of energy condition.  There are several different ways to formulate the energy conditions, but in this work we consider only (i) the Null energy condition (NEC), (ii) Weak energy condition (WEC), and
(iii) Strong energy condition (SEC), which reads through the following inequalities
 \begin{subequations}
\label{48}
\begin{align}
NEC &:& \rho (r)+  p \geq 0, \\
WEC&:& \rho + p \geq 0, ~~ \text{and}~~~\rho(r)+ \frac{q^2}{8 \pi r^4} \geq 0, \\
SEC &:& \rho + p \geq 0, ~~ \text{and}~~~ \rho+3\,p +\frac{q^2}{4 \pi r^4}\geq  0 .
\end{align}
\end{subequations}

\begin{figure}[hbt!] 
\centering
\includegraphics[width=7.5cm]{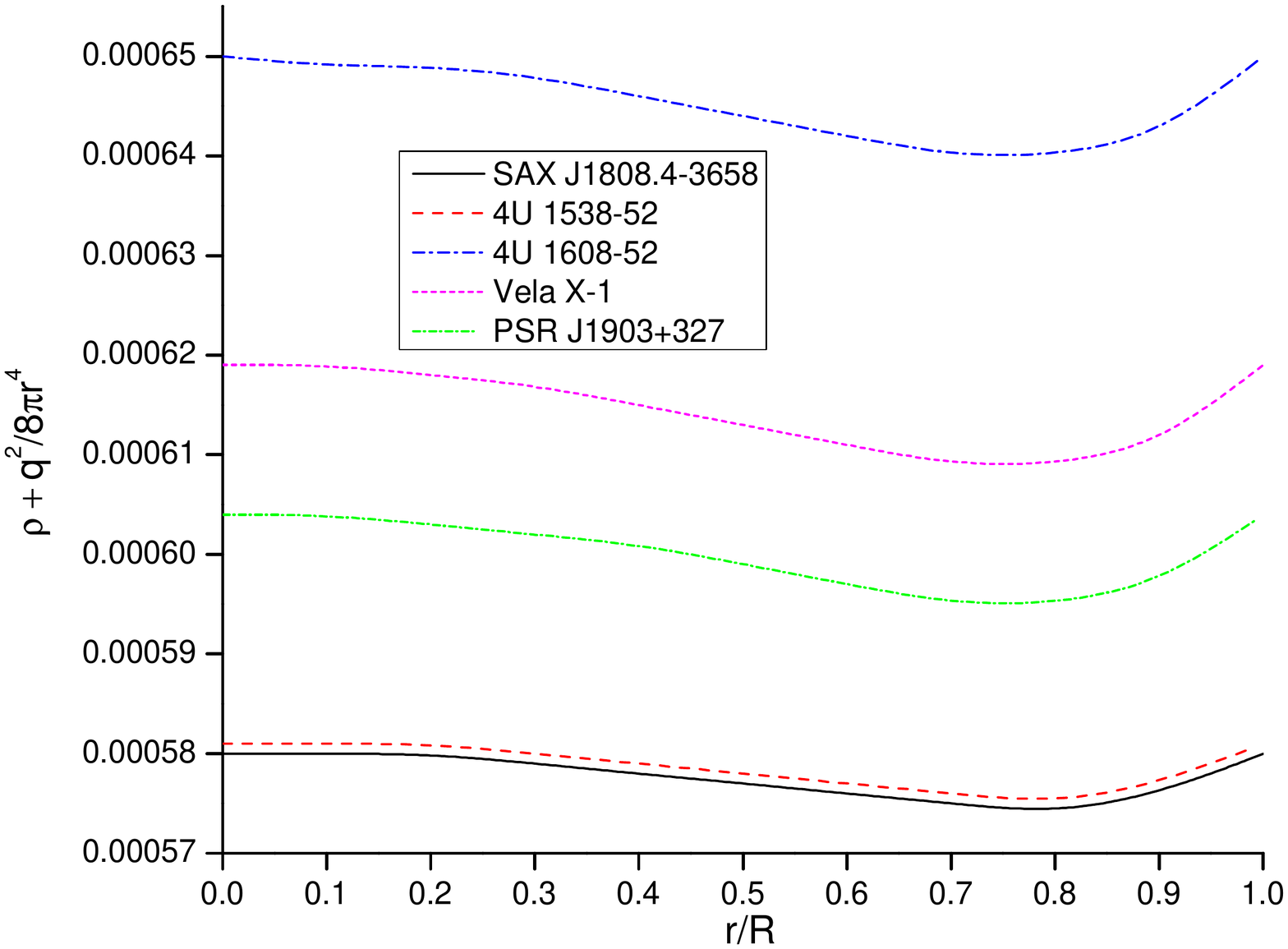} \includegraphics[width=6cm]{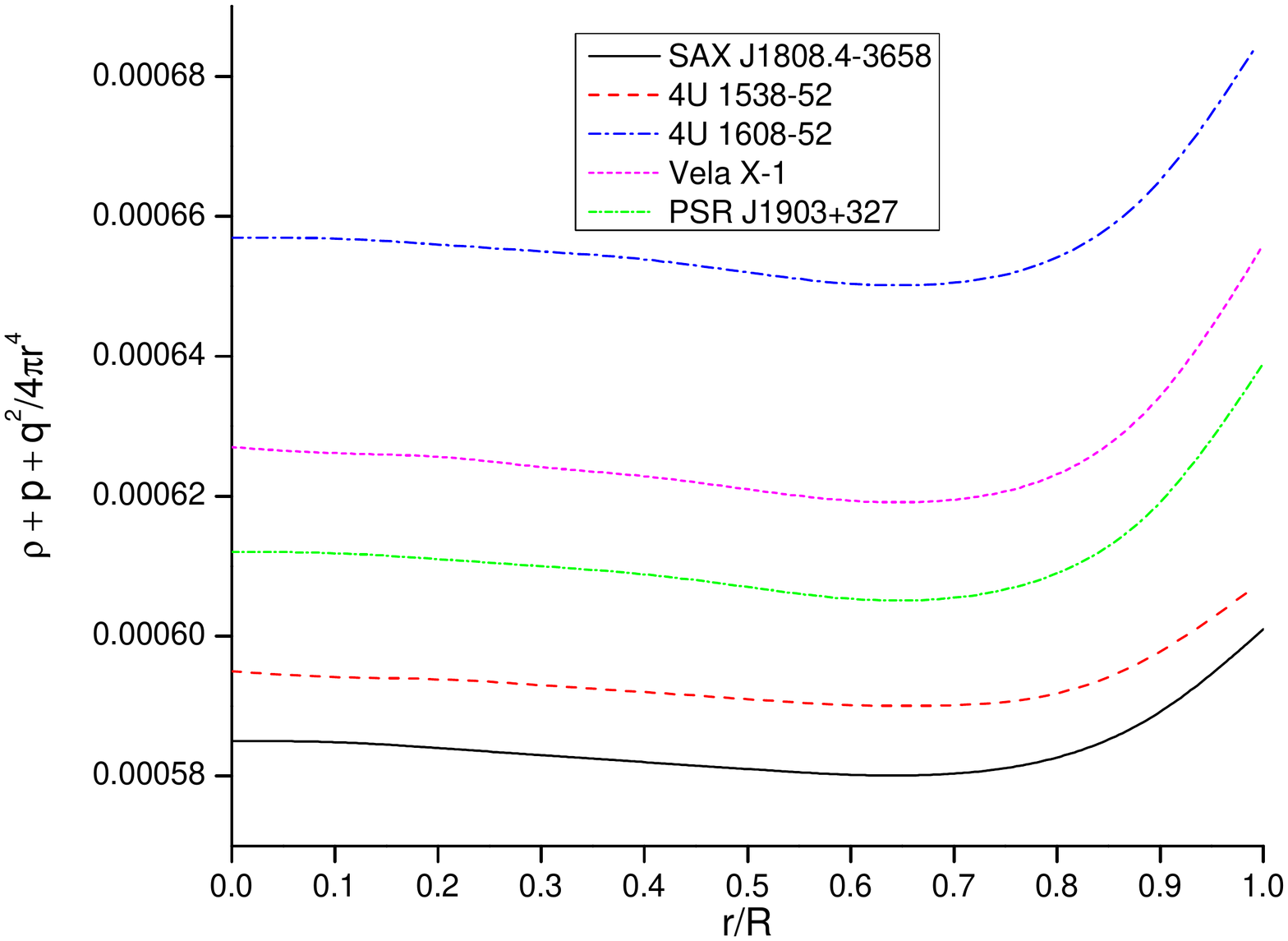}
\includegraphics[width=7cm]{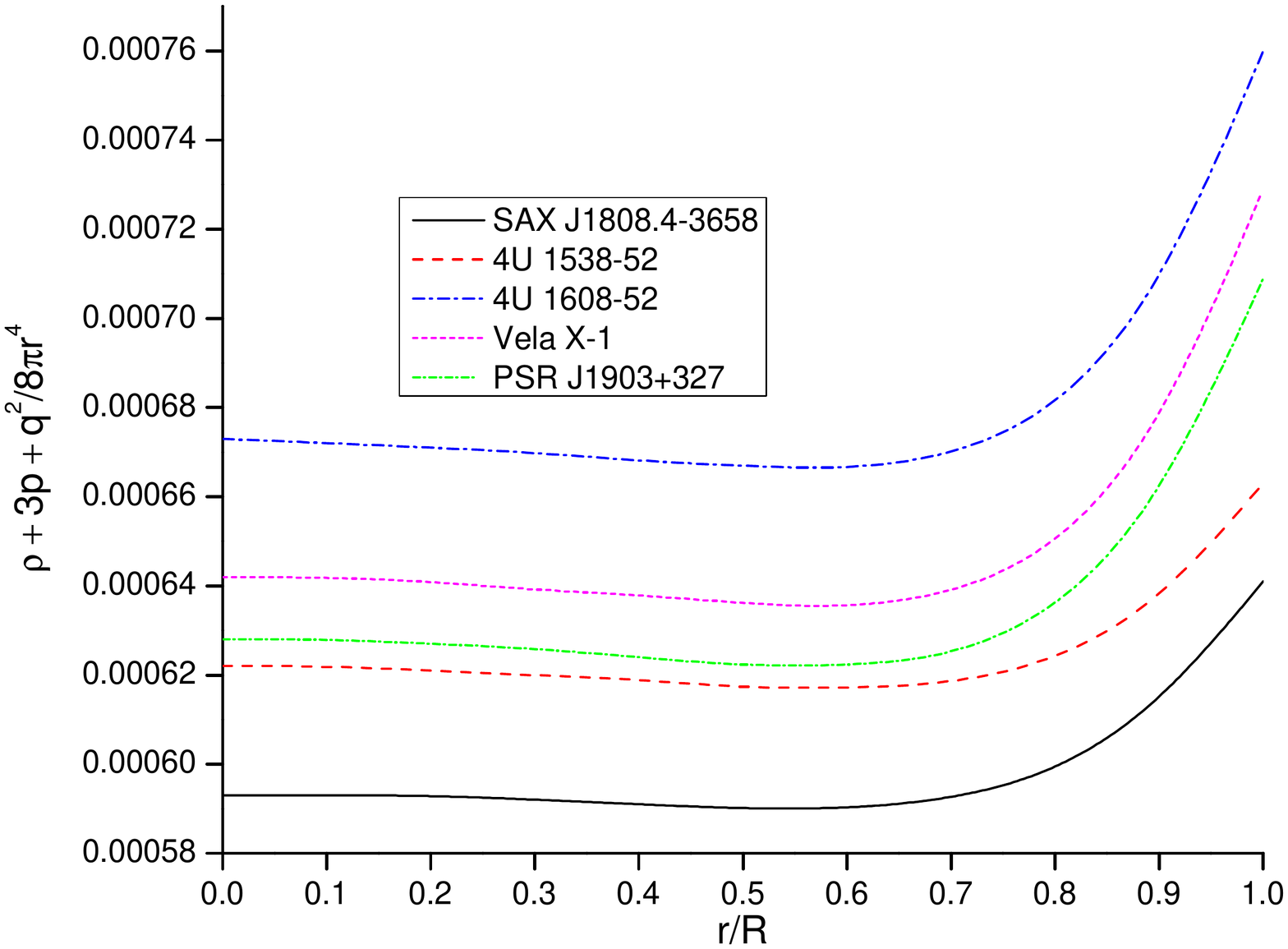}
\caption{\emph{The variation of the NEC, WEC, and SEC for the compact objects SAX J1808.4-3658, 4U 1538-52 and Her X-1. For the purposes of this calculation, we use the same values as given in Fig.\,\ref{f1}.}}
\label{f4}
\end{figure}

To satisfy energy conditions, our system should obey all
the inequalities (\ref{48}), simultaneously. Taking into account the conditions and plotting the left sides of (\ref{48}), one can easily justify the nature of energy conditions as shown in Fig.\,\ref{f4}.
As mentioned above, that classical forms of matter are believed to
obey the energy conditions and our model satisfy all conditions throughout the spacetime.

\begin{table}
\centering \caption{Structural properties of strange star "4U1608-52" within radius}
\scalebox{0.8}{%
\begin{tabular}{c|c|c|c|c|c|c}
\hline \textbf{Radius}  & \multicolumn{6}{c}{Values of the physical parameters $  K=0.00017, \quad CR^2= -0.00008, \alpha^{2}=20.527$}  \\ \hline
x=$\frac{r}{R}$ & Pressure (P) & Density (D) & 	Charge  ($q_1$)  &  $dp/c^{2}d\rho$ &	 P/D  & Redshift ($Z_S$) \\ \hline
0&	0.031855&	1.411525&		0&	0.881124&	0.022568&	1.33224\\ \hline
0.1&	0.031445&	1.411058&		0.00411&	0.873786&	0.022285&	1.322671\\ \hline
0.2&	0.030223&	1.409641&		0.033025&	0.851794&	0.02144&	1.293971\\ \hline
0.3&	0.028207&	1.407223&		0.112279&	0.815206&	0.020045&	1.246165\\ \hline
0.4&	0.025436&	1.403713&		0.268969&	0.764103&	0.018121&	1.179284\\ \hline
0.5&	0.021968&	1.398968&		0.532792&	0.69857&	0.015703&	1.093355\\ \hline
0.6&	0.017891&	1.392773&		0.937502&	0.61865&	0.012846&	0.988377\\ \hline
0.7&	0.013343&	1.384804&		1.523096&	0.524288&	0.009635&	0.86429\\ \hline
0.8&	0.008545&	1.374567&		2.339397&	0.415221&	0.006216&	0.720915\\ \hline
0.9&	0.003869&	1.361264&		3.452403&	0.290812&	0.002842&	0.557856\\ \hline
1&	0&	1.343531&		4.956751&	0.149728&	0&	0.374314\\ \hline
\label{table4}
\end{tabular}}
\end{table}

\begin{table}
\centering \caption{Structural properties of strange star "Vela X-1" within radius}
\scalebox{0.8}{%
\begin{tabular}{c|c|c|c|c|c|c}
\hline \textbf{Radius}  & \multicolumn{6}{c}{Values of the physical parameters $ K=0.00019,\quad CR^2= -0.00009, \quad \alpha^{2}=20.506 $}  \\ \hline
x=$\frac{r}{R}$ & Pressure (P) & Density (D) & 	Charge  ($q_1$)  &  $dp/c^{2}d\rho$ &	 P/D  & Redshift ($Z_S$) \\ \hline
0&	0.019158&	1.420783&		0&	0.590517&	0.013484&	1.350406\\ \hline
0.1&	0.018868&	1.420289&		0.004235&	0.584838&	0.013284&	1.340723\\ \hline
0.2&	0.018004&	1.41879&		0.034026&	0.567802&	0.01269&	1.31168\\ \hline
0.3&	0.016588&	1.416233&		0.115696&	0.539409&	0.011713&	1.263289\\ \hline
0.4&	0.014659&	1.412518&		0.277197&	0.499647&	0.010378&	1.195565\\ \hline
0.5&	0.012278&	1.407492&		0.549204&	0.448466&	0.008723&	1.108509\\ \hline
0.6&	0.009541&	1.400924&		0.966644&	0.385742&	0.00681&	1.002092\\ \hline
0.7&	0.006598&	1.392466&		1.570997&	0.311211&	0.004738&	0.876217\\ \hline
0.8&	0.003691&	1.381579&		2.414076&	0.22437&	0.002672&	0.730665\\ \hline
0.9&	0.001236&	1.367403&		3.56476&	0.124295&	0.000904&	0.564992\\ \hline
1&	0&	1.34845&		5.122273&	0.009315&	0&	0.378343 \\ \hline
\label{table5}
\end{tabular}}
\end{table}

\begin{table}
\centering \caption{Structural properties of strange star "PSR J1903+327" within radius}
\scalebox{0.8}{%
\begin{tabular}{c|c|c|c|c|c|c}
\hline \textbf{Radius}  & \multicolumn{6}{c}{Values of the physical parameters $ K=0.0002,  CR^2 =-0.000091, \alpha^{2}= 20.18$} \\ \hline
x=$\frac{r}{R}$ & Pressure (P) & Density (D) & 	Charge  ($q_1$)  &  $dp/c^{2}d\rho$ &	 P/D  & Redshift ($Z_S$)  \\ \hline
0&	0.01616&	1.364727&		0&	0.540955&	0.011841&	1.246348\\ \hline
0.1&	0.015913&	1.364268&		0.004007&	0.535778&	0.011664&	1.237463\\ \hline
0.2&	0.015178&	1.362876&		0.03219&	0.520244&	0.011136&	1.210813\\ \hline
0.3&	0.013974&	1.360502&		0.109421&	0.494347&	0.010271&	1.166407\\ \hline
0.4&	0.012335&	1.35706&		0.262049&	0.458059&	0.009089&	1.104254\\ \hline
0.5&	0.010316&	1.352413&		0.518881&	0.411319&	0.007628&	1.024355\\ \hline
0.6&	0.008&	1.346357&		0.912533&	0.353992&	0.005942&	0.926681\\ \hline
0.7&	0.005519&	1.33859&		1.481438&	0.28582&	0.004123&	0.81115\\ \hline
0.8&	0.00308&	1.328648&		2.273085&	0.206329&	0.002318&	0.677576\\ \hline
0.9&	0.00103&	1.315804&		3.349655&	0.114684&	0.000783&	0.525593\\ \hline
1&	0&	1.298829&		4.798805&	0.009419&	0&	0.354509 \\ \hline
\label{table6}
\end{tabular}}
\end{table}

\section{Final remarks}
\label{fin}
Compact objects have a main role in relativistic astrophysics for several reasons.
For instance neutron stars are the most stable
compact objects in the universe, but the maximal mass value of such objects is still
an open question to the researcher. In this work we have considered the Buchdahl ansatz \cite{Buchdahl} to
find an exact solution for the stability limit of relativistic charged spheres in the context of Einstein-Maxwell theory. This would specifically be the case for the electric fields carried by hypothetical compact stars made of perfect fluid matter, and the exterior spacetime is represented by the Reissner-Nordstrom metric. Our motivation is to explore a class of exotic astrophysical objects with similar mass and radii, like SAX J1808.4-3658, 4U 1538-52, PSR J1903+327 etc, which are confirmed by observations of gamma-ray repeaters and anomalous X-ray pulsars. The novelty of our approach is that the charged fluid sphere has been constructed form Buchdahl ansatz by using a suitable transformation to solve the system of hypergeometric equations. The parameter $K$ plays an important role in determining the stellar structure. 
An important feature of this ansatz is that the energy density and pressure are maximum at the center and monotonically decreasing towards the boundary which ensures the singularity free stellar structure. The main result of this work depends on the transformation of Eq. (\ref{10}).  Further motivation for this ansatz was drawn from astrophysical considerations that charged sphere Coulomb repulsion tends to oppose the gravitational force, $R_c$ should be less than ($9/4) M$.  By varying the fundamental parameters of each model, we analyzed some limits found in GR, such as the modified TOV equation, the Buchdahl bound, and the velocity of sound which ensures the required physical criteria for stable stellar structure. The analysis of this problem is important, because, we show that isotropic EoS is a best and reasonable hypothesis to explain the compactness of the charged star.

Based on physical requirements, the interior solution connects smoothly at $r = R$, with the exterior solution given by Reissner-Nordstr$\dot{\text{o}}$m metric. Then we have compared the both side metrics, and constants are determined which are enlisted in Table \textbf{I}. Regarding the physical properties we saw that energy density is always take greater values throughout the stellar interior as we can see from Fig.\,\ref{f1}. We have studied the effects of electric field intensity $q(r )$, which is continuous in the interior and vanishes at the centre \cite{Komathiraj}. Moreover, electric field intensity is positive and monotonically increasing. For a better visualization of the results shown in Fig.\,\ref{f1}. In a strong gravitational field, Ray \emph{et al} \cite{Ray1} showed that to see any appreciable effect on compact stars the electric fields have to be huge, approximately $10^{21} V/m$. It is important to note that increase in mass is primarily brought in by the softening of the pressure gradient because of the Coulombian term coupled with the Gravitational matter part. We would also like to mention here that from our model as the radius of the star increases, the electric charge is also increasing, which means  the gravitational attraction is counterbalanced by the repulsive Coulombian force. The global balance of the forces allows a huge charge ($\sim 10^{20}$  Coulomb) to be
 present in a neutron star producing a very high electric field ($ \sim 10^{21}V/m$) \cite{Varela,SRay,SRay1,Ghezzi}.
 The charged stars have large mass and radius as we should expect due to the effect of the repulsive Coulomb force with the $M/R$ ratio increasing with charge. In the limit of the maximum charge the mass goes up to $ \sim $ 10$ M_{\odot} $ which is much higher than the maximum mass allowed for a neutral compact star. However, the local effect
 of the forces experienced by a single charged particle, makes it to discharge quickly. This creates a global force imbalance and the system collapses to a charged black hole.

Mainly, we perform a detailed investigation of the physical implications of high density system like a charged perfect fluid, and checked the physical viability and acceptability of the present model in connection with a number of compact star candidates like SAX J1808.4-3658, 4U 1538-52, PSR J1903+327, Vela X-1 and 4U1608-52. We have analyzed the configurations for several values of constant parameters depending on $\alpha$, $C$ and $K$. The permitted
values of the unknown parameters are determined from the matching conditions at the boundary
and pressure at the boundary is zero i.e., $p(R)$ = 0. Then using these values we perform the
mass-radius relation for each compact objects, as evident in Table~ \ref{table1}.
The consequence of mass-radius relation is that for an isotropic spheres with constant density
must be less than $8/9$ \cite{Buchdahl}, which is satisfied for every compact objects, as seen form Table~ \ref{table1}.

For clarity, we explore several aspects of the model stating form energy conditions,
velocity of sound and the stability of the system using modified TOV equation. We analysed every
cases step by step with graphical display in order to verify the model that can be considered viable
within the specified observational constraint. Considering observed masses of the compact objects namely,
SAX J1808.4-3658, 4U 1538-52, PSR J1903+327, Vela X-1 and 4U1608-52, we explore the interior of the
star. The results are summarized in Table~ \ref{table2}-\ref{table6}, which
strongly suggest that a class of compact stellar models with charged perfect fluid
matter distribution are permitted with the new solution discussed here.  In this work, the density  $\rho = \frac{D\,c^2}{8\pi\,G\,R^2} ~ g/cm^{3}$ see Table~\ref{table1}, the pressure  $ p=\frac {P\,c^4}{8\pi\,G\,R^2}~dyne/cm^{2} $ and the charge $ q=\frac{\sqrt{q_1}~ r^2}{R} cm$, where $G= 6.673 \times 10^{-8}\, cm^3/g s^2$, $c=2.997\times10^{10} cm/s$, $R = radius (in \, cm)$.

Finally we hope that the procedure which we describe in this article, will lead to many other interesting possibility in the future, and thereby describe some other types of compact objects should also be investigated. We shall report such
analyses in a future work.

\section{acknowledgments}
SKM acknowledges support from the authority of University of Nizwa, Nizwa, Sultanate of Oman. AB thanks the organizers of University of Kwazulu-Natal for financial support.

\appendix
\section{Solution of Differential Equations}
We will follow (\ref{15}) to show that, how one can solve the 2nd order differential equation.
For the Eq. (\ref{15}), we compare with
\begin{eqnarray}
W_{0}\frac{d^{2}Y}{dX^{2}}+W_{1}\frac{dY}{dX}+W_{2}X=R  \label{a1}
\end{eqnarray}
where $ W_{0}= X^{2}\alpha^{2}+X,~~~ W_{1}= 0,~~~~  W_{2}= -2\alpha^{2},~~~~  R= 0.$ For exactness one can verify that $ W_{2}-\frac{dW_{1}}{dx}+\frac{d^{2}W_{0}}{dx^{2}}=0 $, which shows that  the equation (\ref{15}) is exact. Hence the primitive of the given differential equation is
\begin{center}
$ W_{0}\frac{dY}{dX}+(W_{1}-W'_{0})X=\int R dX +A$,
\begin{eqnarray}
\frac{dY}{dX}-\frac{2X\alpha^{2}+1}{X^{2}\alpha^{2}+X}Y=\frac{A}{X^{2}\alpha^{2}+X}\label{ab}
\end{eqnarray}
\end{center}
where $ A $ is arbitrary constant. Now, solving the
above the Eq. (\ref{ab}), which yield
\begin{eqnarray}
\resizebox{0.95\hsize}{!}{$ Y(X)=(X^{2}\alpha^{2}+X)\bigg[A\alpha^{2}\bigg(\sin^{2}(\arctan(\sqrt{X\alpha^{2}})) -2\csc^{2}(\arctan(\sqrt{X\alpha^{2}}))-\log(\sin^{2}(\arctan(\sqrt{X\alpha^{2}}))\bigg)+B ~\bigg]$},~~\label{190}
\end{eqnarray}
where $A$ and $B$ are constants.

\end{document}